\documentclass[
  aps,
  prb,
  showpacs,
  superscriptaddress,
  floatfix,
  twocolumn,
  secnumarabic,
  amssymb,
  amsmath]{revtex4-1}

\usepackage{graphicx}
\usepackage{bm}
\usepackage{color}
\usepackage[colorlinks=true,citecolor=blue,linkcolor=blue,pdfstartview=FitH,pdfauthor=Gepraegs]{hyperref}

\begin{document}

\preprint{Gepraegs \textit{et al.}, version: 2013-04-23}

\title{Converse Magnetoelectric Effects in Fe$_{3}$O$_{4}$/BaTiO$_{3}$ Multiferroic Hybrids}

\author{Stephan~Gepr\"{a}gs}
 \email{Stephan.Gepraegs@wmi.badw.de}
 \affiliation{Walther-Mei{\ss}ner-Institut, Bayerische Akademie der
              Wissenschaften, 85748 Garching, Germany}

\author{Dan~Mannix}
 \affiliation{Institut N\'{e}el, CNRS-UJF, 38042 Grenoble Cedex 9, France }

\author{Matthias~Opel}
 \affiliation{Walther-Mei{\ss}ner-Institut, Bayerische Akademie der
              Wissenschaften, 85748 Garching, Germany}

\author{Sebastian~T.~B.~Goennenwein}
 \affiliation{Walther-Mei{\ss}ner-Institut, Bayerische Akademie der
              Wissenschaften, 85748 Garching, Germany}

\author{Rudolf~Gross}
 \affiliation{Walther-Mei{\ss}ner-Institut, Bayerische Akademie der
              Wissenschaften, 85748 Garching, Germany}
 \affiliation{Physik-Department, Technische Universit\"{a}t M\"{u}nchen, 85748 Garching, Germany}

\date{\today}%

\begin{abstract}
The quantitative understanding of converse magnetoelectric effects,~i.e., the variation of the magnetization as a function of an applied electric field, in extrinsic multiferroic hybrids is a key prerequisite for the development of future spintronic devices. We present a detailed study of the strain-mediated converse magnetoelectric effect in ferrimagnetic Fe$_{3}$O$_{4}$ thin films on ferroelectric BaTiO$_{3}$ substrates at room temperature. The experimental results are in excellent agreement with numerical simulation based on a two-region model.  This demonstrates that the electric field induced changes of the magnetic state in the Fe$_{3}$O$_{4}$ thin film can be well described by the presence of two different ferroelastic domains in the BaTiO$_{3}$ substrate, resulting in two differently strained regions in the Fe$_{3}$O$_{4}$ film with different magnetic properties. The two-region model allows to predict the converse magnetoelectric effects in multiferroic hybrid structures consisting of ferromagnetic thin films on ferroelastic substrates.
\end{abstract}


\maketitle

\section{Introduction}
\label{sec:intro}

The manipulation of the magnetization by electric fields, which is referred to as converse magnetoelectric effect, is one of the most challenging tasks in today's spintronic devices.\cite{Binek:17:2005,Eerenstein:442:2006,Ramesh:6:2007} The presence of such a cross-coupling between ordered magnetic and dielectric states in so-called multiferroic systems allows for the development of a new family of devices, where the spin degree of freedom is controlled by electric instead of magnetic fields.\cite{Goennenwein:2:2008,Weiler:11:2009,Brandlmaier:110:2011,Hu:2:2011,Heron:107:2011,Hu:24:2012,Vaz:24:2012} It has been shown that a robust electric field control of magnetism can be realized by using extrinsic multiferroic hybrid structures consisting of ferroelectric and ferromagnetic materials.\cite{Gepraegs:87:2007,Nan:103:2008,Vaz:22:2010,Ma:AdvMater:23:2011,Brandlmaier:85:2012} In these systems, the extrinsic magnetoelectric effects rely on either electric field effects\cite{Chiba:10:2011} using carrier mediated ferromagnets,\cite{Molegraaf:21:2009} or employing exchange coupling effects between antiferromagnetic, ferroelectric and ferromagnetic compounds.\cite{Hochstrat:272:2004,Chu:7:2008} Moreover, from the beginning the elastic coupling at the interface between ferromagnetic and ferroelectric materials has been exploited to realize a robust electric field control of the magnetization.\cite{Suchtelen:27:1972,Thiele:75:2007,Eerenstein:6:2007} In this case, the electric and magnetic degrees of freedom are elastically coupled at the interface between the two ferroic constituents. More precisely, strain mediated converse magnetoelectric effects rely on an electric field induced strain in the ferroelectric via the converse piezoelectric effect, or via ferroelastic reorientations. This strain is then transferred into the ferromagnetic thin film clamped onto the ferroelectric, where converse magnetoelastic effects modify the magnetization.

Indirect strain-mediated magnetoelectric effects can be realized using particulate composites,\cite{Run:9:1974} laminated composites,\cite{Ryu:40:2001} nanostructured composites,\cite{Zheng:303:2004} and horizontal multiferroic hybrid structures. In the latter case, the discovery of giant sharp and persistent converse magnetoelectric effects in ferromagnetic thin films epitaxially grown on BaTiO$_{3}$ (BTO) crystals\cite{Eerenstein:6:2007} triggered an increasing research effort in this field, using different ferromagnetic material systems, such as magnetic ferrites (Fe$_{3}$O$_{4}$,\cite{Vaz:94:2009,Ziese:88:2006,Tian:92:2008,Sterbinsky:96:2010,Koo:94:2009,Niranjan:PRB:78:2008} CoFe$_{2}$O$_{4}$,\cite{Chopdekar:89:2006,Pan:46:2013} and NiFe$_{2}$O$_{4}$\cite{Chopdekar:86:2012}), magnetic perovskites and double-perovskites (La$_{1-x}$Sr$_{x}$MnO$_{3}$,\cite{Dale:82:2003} La$_{1-x}$Ca$_{x}$MnO$_{3}$,\cite{Singh:99:2006,Alberca:86:2012} Pr$_{1-x}$Ca$_{x}$MnO$_{3}$,\cite{Murugavel:85:2004} and Sr$_{2}$CrReO$_{6}$\cite{Czeschka:95:2009,Komelj:82:2010,Opel:208:2011}), and $3d$-magnets (Fe,\cite{Shirahata:99:2011,Venkataiah:111:2012,Sahoo:76:2007,Venkataiah:99:2011} Ni,\cite{Weiler:11:2009,Gepraegs:96:2010,Shu:100:2012,Ghidini:4:2013,Streubel:87:2013} Co,\cite{Narayanan:12:2012} and CoFe\cite{Lahtinen:2:2012}). In particular, strong electric field-induced changes of the magnetic state were reported in Fe$_{3}$O$_{4}$-based extrinsic multiferroic hybrids.\cite{Brandlmaier:77:2008,Liu:19:2009,Vaz:94:2009,Liu:107:2010} Since Fe$_{3}$O$_{4}$ has a ferrimagnetic ground state with a high Curie temperature of about 850\,K,\cite{OHandley:book} a pronounced magnetostrictive effect,\cite{Arai:34:1976} and a high spin polarization at the Fermi level,\cite{Zhang:44:1991} Fe$_{3}$O$_{4}$ is a promising candidate for possible future magnetoelectric devices. Moreover, a sizable spontaneous ferroelectric polarization was reported in Fe$_{3}$O$_{4}$ films\cite{Alexe:21:2009,Takahashi:86:2012} which is caused by a non-centrosymmetric charge order.\cite{Nazarenko:97:2006,Yamauchi:79:2009} Therefore, Fe$_{3}$O$_{4}$ seems to be not only the oldest magnetic material but also the first magnetoelectric multiferroic known to mankind.\cite{Rado:12:1975}

\begin{figure*}[tbh]
  \includegraphics[width=2\columnwidth]{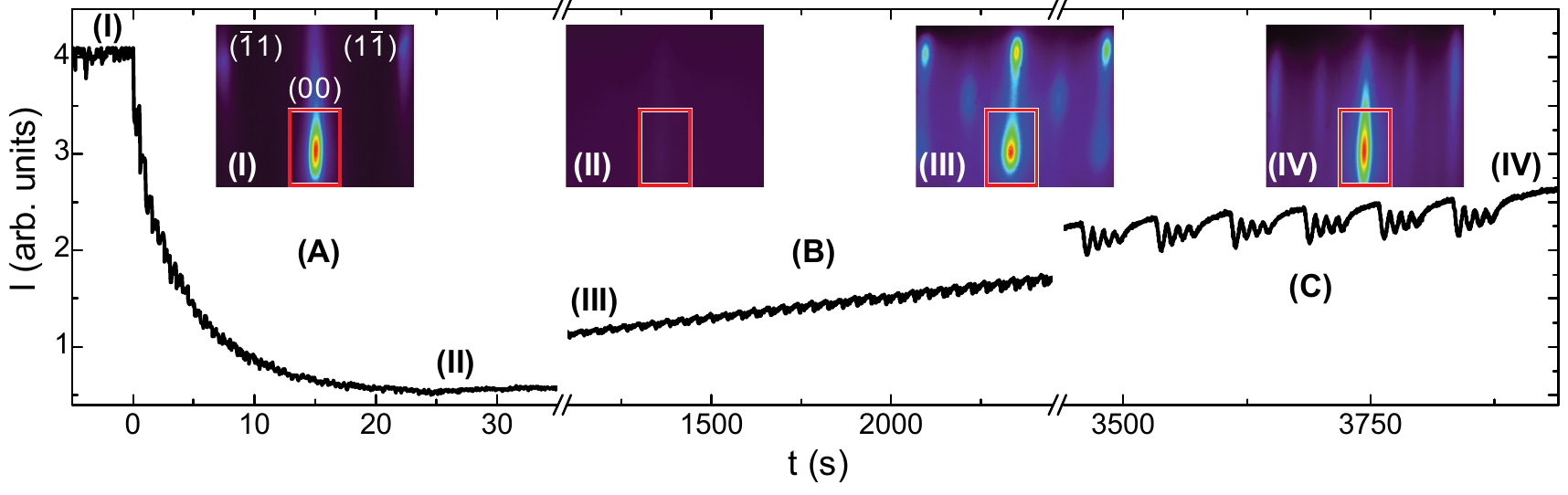}
    \caption{(color online) Intensity evolution of the RHEED $(00)$ reflection monitored during the deposition of a 64\,nm thick Fe$_{3}$O$_{4}$ film on a 0.5\,mm thick BTO substrate. The intensity was integrated within the red rectangle marked in the RHEED patterns, which are shown in the insets. The time stamps at which the RHEED patterns were recorded are marked by (I)--(IV). The growth process can be subdivided into three main steps: highly disordered growth (A) (0\,s $\leq t \lesssim$ 25\,s) --- three dimensional growth mode (B) (300\,s $\lesssim t \lesssim$ 3000\,s) --- layer-by layer growth mode (C) (3000\,s $\lesssim t \lesssim$ 3950\,s).}
    \label{fig:Fig1}
\end{figure*}

To quantitatively understand the strain-mediated magnetoelectric effects in BTO-based multiferroic hybrid structures, as a first step, we recently investigated the magnetization changes as a function of temperature in Fe$_{50}$Co$_{50}$/BTO and Ni/BTO hybrids caused by magnetoelastic effects.\cite{Gepraegs:86:2012} We demonstrated that the manipulation of the magnetization in BTO-based multiferroic hybrid structures can be theoretically well described by controlling the ferromagnetic and ferroelastic domain configuration in the multiferroic hybrid. Here, as a second step, we show that by knowing the ferroelastic domain configuration in the BTO substrate as a function of the applied electric field, the strain-mediated converse magnetoelectric effect in epitaxially grown Fe$_{3}$O$_{4}$/BTO hybrid structures can be theoretically simulated using a simple model, consisting of two magnetic regions in the Fe$_{3}$O$_{4}$ thin film. This demonstrates that by controlling the ferroelastic domain state in the BTO substrate large and robust manipulations of the magnetization as a function of the applied electric field are feasible at room temperature in BTO-based multiferroic hybrid structures.

The article is organized as follows: In Sec.~\ref{sec:exp}, we discuss the growth process and the structural as well as magnetic properties of Fe$_{3}$O$_{4}$ thin films epitaxially grown on BTO substrates. Since X-ray diffraction measurements using synchrotron radiation carried out on these Fe$_{3}$O$_{4}$/BTO hybrid structures reveal two differently strained regions in the Fe$_{3}$O$_{4}$ thin film at low electric fields, a two-region model is introduced in Sec.~\ref{sec:model}. In this model, the room-temperature magnetization behavior of Fe$_{3}$O$_{4}$/BTO hybrids as a function of the applied electric field can be explained by the magnetic behavior of parts of the Fe$_{3}$O$_{4}$ thin film elastically clamped onto two different ferroelastic domains of the BTO substrate. Thus, to simulate the strain-mediated converse magnetoelectric effect, at first the volume fraction of the ferroelastic domains in the BTO substrate is determined (cf.~Sec.~\ref{sec:xc}), and then the magnetic state of the Fe$_{3}$O$_{4}$ thin film on top of these domains is calculated using magnetoelastic theory (cf.~Sec.~\ref{sec:MaMc}). In Sec.~\ref{sec:ExpSimu}, we show that there is an excellent agreement between the experimental results and the theoretical simulations based on the two-region model. This demonstrates that the experimentally obtained room-temperature converse magnetoelectric effects in BTO-based multiferroic hybrids are well described within this approach for magnetic fields larger than the magnetic saturation field.

\section{Thin film growth}
\label{sec:exp}

The Fe$_{3}$O$_{4}$ thin films with thicknesses between 40\,nm and 70\,nm were epitaxially grown on 0.5\,mm thick (001)-oriented BTO substrates by laser molecular-beam epitaxy (laser-MBE)\cite{Gross:4058:2000} monitored by an \textit{in-situ} high pressure RHEED system.\cite{Klein:12:1999,Klein:211:2000} The energy density of the KrF excimer laser ($\lambda=248$\,nm) at the target was set to 3.1\,J/cm$^2$ and the laser repetition rate was 2\,Hz. The deposition was carried out in an atmosphere of pure argon with a pressure of $1.6\times10^{-3}$\,mbar at a temperature of 593\,K, i.e., in the cubic phase of the BTO substrate.

The growth process of the Fe$_{3}$O$_{4}$ thin films is illustrated in Fig.~\ref{fig:Fig1} on the basis of the intensity evolution of the RHEED $(00)$ spot, which was recorded during the growth of a 64\,nm thick Fe$_{3}$O$_{4}$ film. Before starting the growth process, the RHEED pattern of the BTO substrate reveals three spots, which belong to the first Laue circle and can be indexed with $(\overline{1}1)$, $(00)$, and $(1\overline{1})$ [cf.~inset (I) of Fig.~\ref{fig:Fig1}]. By starting the deposition at $t=0$\,s, the high intensity of the $(00)$ spot drastically decreases [cf.~Fig.~\ref{fig:Fig1} (A)]. After 50\,pulses ($t \approx 25$\,s), the deposition was stopped [cf.~Fig.~\ref{fig:Fig1} (II)]. The RHEED pattern recorded at this moment does not show any reflections  [cf.~inset (II) of Fig.~\ref{fig:Fig1}]. This indicates a highly disordered and nearly amorphous growth within the first monolayers of Fe$_{3}$O$_{4}$. Further deposition leads to a three dimensional crystalline growth as indicated by a checkerboard like RHEED pattern [cf.~inset (III) of Fig.~\ref{fig:Fig1}]. Upon further continuing the deposition, the three-dimensional growth mode turns over into a two-dimensional one with increasing RHEED intensity and emerging intensity oscillations [cf.~Fig.~\ref{fig:Fig1} (B)]. In the last stage of the growth process ($t>3000$\,s), a real two-dimensional layer-by-layer growth mode can be obtained [cf.~Fig.~\ref{fig:Fig1} (C)]. At this stage, a single unit cell of Fe$_{3}$O$_{4}$, which is manifested by four RHEED oscillations,\cite{Reisinger:77:2003} was deposited followed by a growth interruption of 30\,s, allowing for the relaxation of the film surface. The RHEED pattern recorded at the end of the growth process [cf.~inset (IV) of Fig.~\ref{fig:Fig1}] does not show any indication of transmission like spots, suggesting a smooth surface of the Fe$_{3}$O$_{4}$ thin film. Using atomic force microscopy as well as X-ray reflectometry, a surface roughness (rms value) ranging between 0.8\,nm and 2.5\,nm was found. Therefore, as obvious from Fig.~\ref{fig:Fig1}, Fe$_{3}$O$_{4}$/BTO hybrid structures can be fabricated in a true layer-by-layer growth mode. However, the highly disordered growth at the beginning of the deposition of Fe$_{3}$O$_{4}$, which might be explained by the lattice mismatch of Fe$_{3}$O$_{4}$ and BTO of 4.7\% at the growth temperature, results in a large mosaic spread. X-ray diffraction measurements (not shown here) reveal no secondary phases as well as a full width at half maximum of the rocking curves around the Fe$_{3}$O$_{4}$\,(004) reflection of about $0.6^\circ$. This is more than one order of magnitude larger than observed for Fe$_{3}$O$_{4}$ thin films grown on lattice-matched MgO substrates.\cite{Venkateshvaran:79:2009}

\begin{figure}[tb]
  \includegraphics[width=\columnwidth]{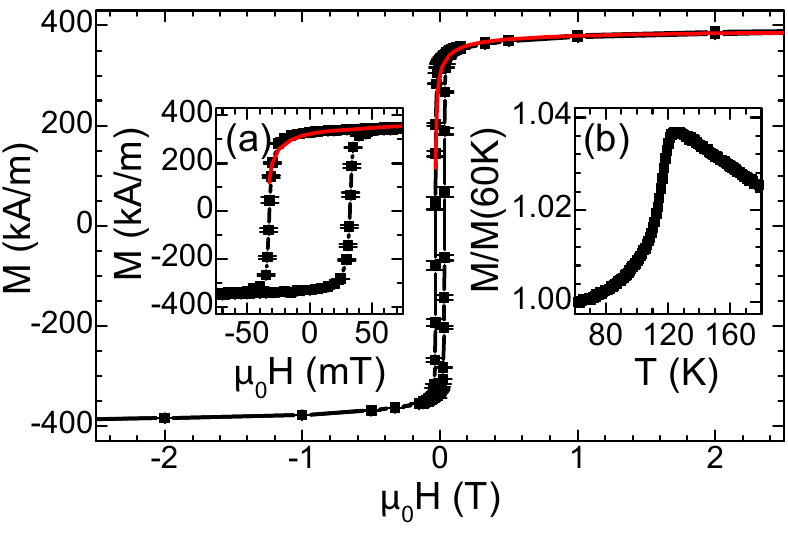}
    \caption{(color online) Magnetic hysteresis of a 45\,nm thick Fe$_{3}$O$_{4}$ film grown on a (001)-oriented BTO substrate measured at 300\,K with the magnetic field applied in the film plane. The magnetic field dependence of the magnetization can be fitted by a $(1-b/\sqrt{H})$ behavior using $b=(0.045\pm0.005)\,\sqrt{\mathrm{T}}$ (red solid line). (a) The $M(H)$ loop reveals a coercive field of 35\,mT and a saturation magnetization of $M_{\mathrm{s}}=438$\,kA/m at $\mu H=7$\,T. (b) The normalized remanent magnetization as a function of temperature measured after field cooling the sample with $\mu_{0}H=7$\,T discloses a Verwey transition at 121\,K.}
    \label{fig:Fig2}
\end{figure}

The magnetic properties of the hybrid structures were studied via superconducting quantum interference device (SQUID) magnetometry in the temperature range between 2\,K and 300\,K at magnetic fields of up to 7\,T applied in the film plane. Figure~\ref{fig:Fig2} shows the magnetic behavior of a 45\,nm thick Fe$_{3}$O$_{4}$ film grown on a BTO substrate. The magnetic hysteresis loop $M(H)$ measured at 300\,K reveals a coercive field of 35\,mT [cf.~inset (a) of Fig.~\ref{fig:Fig2}], which is close to the value reported for Fe$_{3}$O$_{4}$ thin films on lattice-matched MgO substrates.\cite{Reisinger:85:2004} Furthermore, a high saturation magnetization of $M_{\mathrm{s}}=438$\,kA/m at $\mu_{0} H=7$\,T is obtained, which corresponds to 3.1\,$\mu_{\mathrm{B}}/$f.u.. The difference $\Delta M_{\mathrm{s}}$ between $M_{\mathrm{s}}$ and the bulk value $M_{\mathrm{s}}^{\mathrm{bulk}}=4.1\,\mu_{\mathrm{B}}/$f.u. is generally explained by the presence of anti-phase boundaries (APB). APBs exhibit a strong antiferromagnetic coupling, which results in a $(1-b/\sqrt{H})$ behavior of the magnetization, where the parameter $b$ depends on the density of APBs.\cite{Bataille:74:2006} Using $b=(0.045\pm0.005)\,\sqrt{\mathrm{T}}$, the magnetic hysteresis can be nicely fitted [cf.~red line in Fig.~\ref{fig:Fig2}]. This value is much lower than values reported for Fe$_{3}$O$_{4}$ thin films grown on (001)-oriented MgO or Al$_{2}$O$_{3}$ substrates.\cite{Bataille:74:2006,Sofin:83:2011} This might be explained by the disordered growth within the initial phase of the deposition process, leading to a disconnection of the Fe$_{3}$O$_{4}$ thin film with the surface structure of the BTO substrate. Thus, the deviation in saturation magnetization $\Delta M_{\mathrm{s}}$ might not be explained by APBs alone. Oxygen non-stoichiometry Fe$_{3(1-\delta)}$O$_{4}$ with $\delta \neq 0$ is another possibility, which affects $M_{\mathrm{s}}$.\cite{Kakol:40:1989} The degree of oxygen non-stoichiometry $\delta$ can be investigated by measuring the Verwey transition $T_{\mathrm{V}}$, which is a hallmark of the quality of Fe$_{3}$O$_{4}$ thin films. The inset (b) of Fig.~\ref{fig:Fig2} shows the remanent magnetization as a function of temperature measured after field cooling the sample in an external magnetic field of $\mu_{0}H=7$\,T. The Verwey transition is clearly observable at 121\,K, which is identical to the bulk value. Thus, our Fe$_{3}$O$_{4}$ thin films grown on BTO substrates have a low amount of oxygen vacancies and the correct iron oxide phase. Other iron oxides such as maghemite can be safely excluded. Note that a small reduction of the phase transition temperature was observed in other samples, which can be translated into an oxygen non-stoichiometry of $\delta=0.0021$ assuming a linear dependence of $T_{\mathrm{V}}$ on $\delta$.\cite{Shepherd:43:1991} Altogether, the measured reduction of the saturation magnetization with respect to the bulk value might be explained by a combination of effects caused by APBs and a slight oxygen non-stoichiometry in the Fe$_{3}$O$_{4}$ thin films.

\section{Two region model}
\label{sec:model}

One of the intriguing properties of multiferroic hybrid structures consisting of ferromagnetic and ferroelectric materials is the possibility to manipulate the magnetization $\mathbf{M}$ by an electric field $\mathbf{E}$. To investigate this converse magnetoelectric effect in the Fe$_{3}$O$_{4}$/BTO hybrid structures, an Au bottom electrode was sputtered on the backside of each BTO substrate, which enables us to apply an electric field $E_{z}$ across the BTO substrate along the $z$ direction using Fe$_{3}$O$_{4}$ as top electrode. The Fe$_{3}$O$_{4}$/BTO hybrids are further heated to 450\,K well above the Curie temperature of BTO and slowly cooled down to room temperature while applying an electric field of 400\,kV/m.

\begin{figure}[tb]
  \includegraphics[width=\columnwidth]{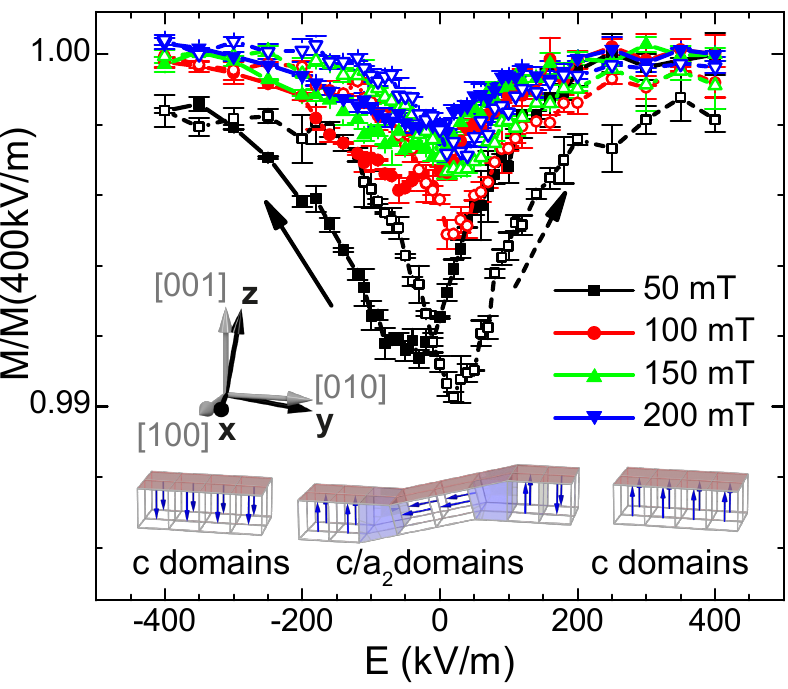}
    \caption{(color online) Converse magnetoelectric effects measured at 300\,K for a Fe$_{3}$O$_{4}$(38\,nm)/BTO multiferroic hybrid. The magnetic field was applied in the film plane along the $y$ direction. A miscut BTO substrate was used for the experiment with a misalignment of $1^\circ$ ($7^\circ$) of the [001] ([100]) direction of the ferroelastic unit cell with respect to the $z$ ($x$) direction describing the surface of the BTO substrate. The variation of the ferroelastic domain configuration as a function of the electric field is illustrated in the bottom part of the figure. At high positive and negative fields, a ferroelastic single $c$ domain state is expected, while at low electric fields a $c$/$a_{2}$ multi-domain state is present in the BTO crystal. The blue arrows indicate the orientation of the ferroelectric polarization and the blue regions denote ferroelastic domain walls. Polarization rotation induced by the electric field is neglected for simplicity.}
    \label{fig:Fig3}
\end{figure}

As an example, Fig.~\ref{fig:Fig3} shows the modification of the magnetization as a function of the applied electric field, $M(E)$, for different values of the applied magnetic field. The curves were measured at room temperature,~i.e.,~in the tetragonal phase of BTO, using a Fe$_{3}$O$_{4}$(38\,nm)/BTO multiferroic hybrid. Here, the projection of the magnetization $M$ on the magnetic field direction, which was oriented along the $y$ direction, is plotted as a function of the electric field strength $E_{z}$ applied across the BTO substrate. In analogy to Ni/BTO hybrid structures discussed in Ref.~\onlinecite{Gepraegs:96:2010}, a butterfly shape of the $M(E)$-loops is observable. On the basis of the measured magnetization changes $\Delta M$, a pseudo-linear magnetoelectric coupling coefficient can be estimated using $\alpha_{23}=d_{\mathrm{FM}}/(d_{\mathrm{FM}}+d_{\mathrm{FE}}) \mu_{0} \Delta M / \Delta E$, which results in $\alpha_{23}=3 \times 10^{-15}$\,s/m. This value is lower than the magnetoelectric constant of the prototype intrinsic magnetoelectric material Cr$_{2}$O$_{3}$ ($\alpha=4.13 \times 10^{-12}$\,s/m),\cite{Astrov:11:1960} demonstrating that the magnitude of the converse linear magnetoelectric effect in BTO-based hybrid structures is comparable or even less than in intrinsic magnetoelectric materials due to the unfavorable ratio between the thickness of the ferromagnetic thin film ($d_{\mathrm{FM}}$) and the thickness of the BTO substrate ($d_{\mathrm{FE}}$). However, the observed magnetoelectric effect is robust at room temperature. Only very recently, large room-temperature magnetoelectric effects were reported in magnetoelectric hexaferrites.\cite{Kitagawa:9:2010,Chun:108:2012} This was the first observation of significant magnetoelectric effects in intrinsic magnetoelectric materials at room temperature.

The variation of the magnetization as a function of an applied electric field in BTO based multiferroic hybrids at 300\,K can be explained by taking into account the ferroelastic domains of BTO.\cite{Gepraegs:96:2010} In the tetragonal phase, BTO exhibits three types of ferroelastic domains with the polarization vector pointing either along the out-of-plane direction ($c$ domains) or one of the two in-plane directions ($a_{1}$ domains and $a_{2}$ domains). The tetragonal unit cell of the $a$ domains is oriented with its longer axis parallel to the BTO surface. It induces tensile strain in the overlying ferromagnetic thin film, which modifies the magnetic anisotropy due to converse magnetoelastic effects. Thus, to quantitatively describe converse magnetoelectric effects in BTO-based multiferroic hybrids, the volume fraction of the different ferroelastic domains has to be known. The requirements are relaxed by using miscut BTO crystals. In this case only a single type of $a$ domains is formed.

\begin{figure}[tb]
  \includegraphics[width=\columnwidth]{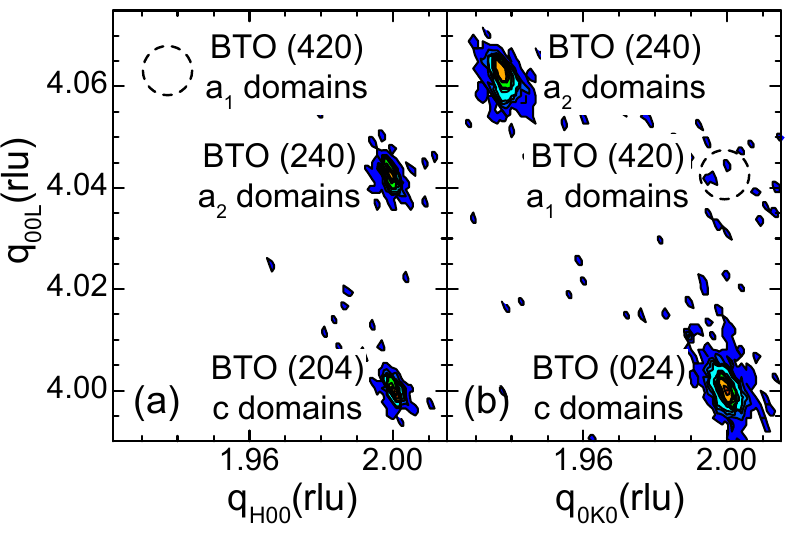}
    \caption{(color online)
X-ray diffraction around the asymmetric BTO (204) reflection of a Fe$_{3}$O$_{4}$(38\,nm)/BTO hybrid recorded in the (a) $q_{H00}-q_{00L}$-plane and (b) $q_{0K0}-q_{00L}$-plane at 0\,kV/m using a miscut BTO crystal (yellow/red: high intensity, light blue/dark blue: low intensity). Only reflections caused by ferroelastic $c$ domains and tilted $a_{2}$ domains are visible. The expected positions of the reflections caused by diffraction from tilted ferroelastic $a_{1}$ domains are marked by dashed circles.}
    \label{fig:Fig4}
\end{figure}

Figure \ref{fig:Fig4} shows x-ray diffraction measurements around the asymmetric BTO~(204) reflection. In the experiment a Fe$_{3}$O$_{4}$(38\,nm)/BTO hybrid with a BTO substrate is used, which has a misalignment of about $1^\circ$ ($7^\circ$) of the [001] ([100]) direction of the ferroelastic unit cells with respect to the $z$ ($x$) direction describing the surface of the BTO crystal (cf.~inset of Fig.~\ref{fig:Fig3}). The mesh scan around the BTO~(204) reflection reveals only two reflections. One is caused by scattering from ferroelastic $c$ domains at $q_{00L}=4.0$\,rlu and the other by tilted ferroelastic $a_{2}$ domains. The second type of $a$ domains ($a_{1}$ domains) could not be detected, neither in the $q_{H00}-q_{00L}$-plane [cf. Fig.~\ref{fig:Fig4}(a)] nor in the $q_{0K0}-q_{00L}$-plane [cf. Fig.~\ref{fig:Fig4}(b)]. Therefore, only a single type of $a$ domains is present in the used miscut BTO crystals at zero electric field. In this case the evolution of the ferroelastic domains as a function of the applied electric field $E_{z}$ can be described as follows: At high positive and negative electric fields, a ferroelastic single $c$ domain state is present, which changes into a ferroelastic multi-domain state around $E_{z}=0$\,kV/m. This ferroelastic multi-domain state consists of ferroelastic $c$ and tilted $a_{2}$ domains as well as $c$$c$ and $a$$c$ domain walls (cf.~bottom part of Fig.~\ref{fig:Fig3}).

\begin{figure}[tb]
  \includegraphics[width=\columnwidth]{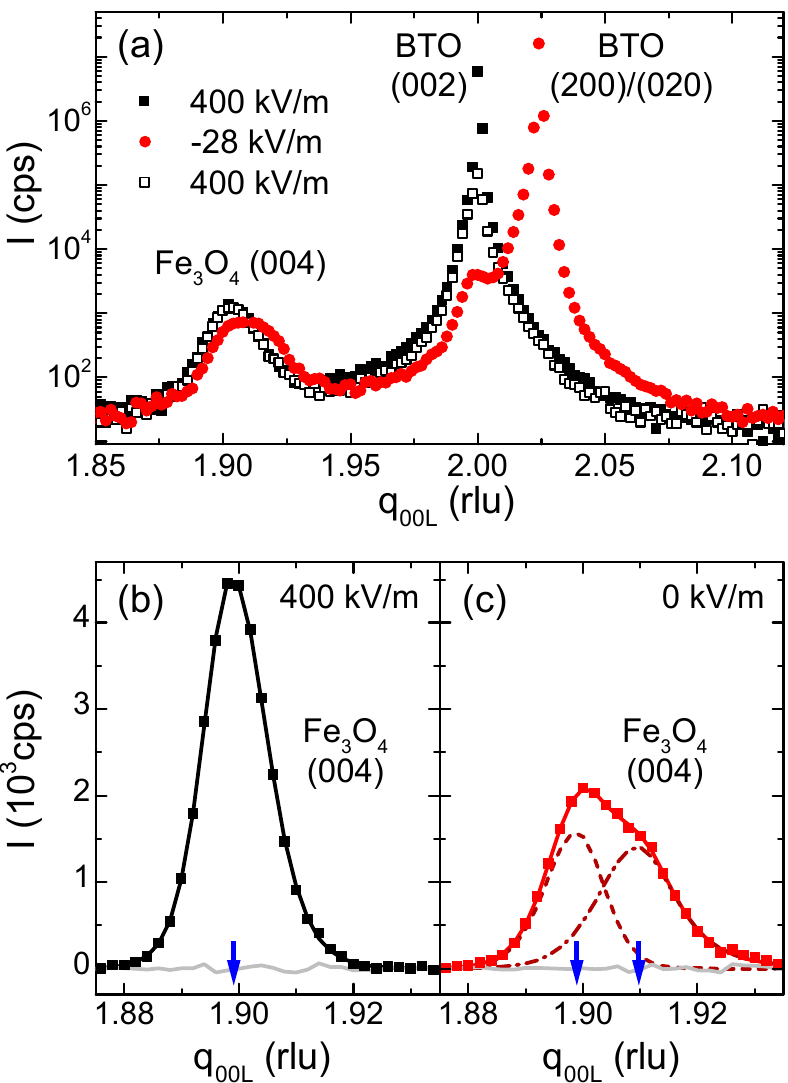}
    \caption{(color online)
(a) X-ray diffraction around the BTO (002) reflection recorded at an electric field of 400\,kV/m (black full symbols) and subsequently of -28\,kV/m (red full symbols) applied across a Fe$_{3}$O$_{4}$/BTO hybrid structure. The measurements were performed at the BM28 beamline at the European Synchrotron Facility (ESRF). Upon increasing the electric field from -28\,kV/m to 400\,kV/m again (black open symbols), the initial measurement at 400\,kV/m is reproduced. The effect of the strain imposed on the Fe$_{3}$O$_{4}$ film is shown in more detail in (b) and (c). In the case of 400\,kV/m, the (004) reflection of the Fe$_{3}$O$_{4}$ thin film (black symbols) can be well fitted using a single peak profile (black solid line). At 0\,kV/m, two peak profiles (dashed lines) are needed to fit the measured reflection (red solid line). The gray lines display the difference between the experimental data (symbols) and the fits (solid lines), and the blue arrows mark the $q_{00L}$ positions of the reflections.}
    \label{fig:Fig5}
\end{figure}

The impact of the ferroelastic domain state on the overlying ferromagnetic thin film due to the imposed strain can be investigated by performing x-ray diffraction measurements on crystalline Fe$_{3}$O$_{4}$/BTO hybrid structures. As an example, Fig.~\ref{fig:Fig5} shows $L$-scans around the BTO~(002) reflection recorded at different electric field strengths $E_{z}$ applied across the Fe$_{3}$O$_{4}$/BTO hybrid structure. The measurements shown in Fig.~\ref{fig:Fig5} were carried out at the BM28 beamline at the European Synchrotron Radiation Facility (ESRF) using synchrotron radiation with an energy of 7.1255\,keV. At $E_{z}=400$\,kV/m, the BTO substrate yields only one reflection at $q_{00L}=2.0$\,rlu [cf.~full black symbols in Fig.~\ref{fig:Fig5}(a)]. This confirms that a single ferroelastic $c$ domain state exists at this electric field strength. By reducing the electric field to values close to the coercive field of BTO ($E_{z}=-28$\,kV/m), a second reflection around $q_{00L}=2.03$\,rlu emerges, which is caused by ferroelastic $a_{2}$ domains formed in the BTO substrate [cf.~full red symbols in Fig.~\ref{fig:Fig5}(a)]. By increasing the electric field to the starting value of $E_{z}=400$\,kV/m, the first measurement is reproduced,~i.e.,~a single ferroelastic $c$ domain state is restored [cf.~open black symbols in Fig.~\ref{fig:Fig5}(a)]. The impact of the strain imposed on the Fe$_{3}$O$_{4}$ thin film was investigated in detail around the Fe$_{3}$O$_{4}$~(004) reflection [Figs.~\ref{fig:Fig5}(b) and (c)]. The peak profile of the Fe$_{3}$O$_{4}$~(004) reflection measured while applying an electric field of $E_{z}=400$\,kV/m can be well fitted using a single Pearson VII peak function\cite{Hall:10:1977} [cf.~solid line in Fig.~\ref{fig:Fig5}(b)]. This demonstrates that the Fe$_{3}$O$_{4}$ thin film is homogeneously strained at this electric field strength. An out-of-plane lattice constant of $c=(0.8492\pm0.0005)$\,nm can be derived from the peak position of the Fe$_{3}$O$_{4}$~(004) reflection. Furthermore, x-ray diffraction measurements carried out around the Fe$_{3}$O$_{4}$~(404) reflection (not shown here) reveal an in-plane lattice constant of $a=(0.8296\pm0.0008)$\,nm. Comparing these values to the bulk lattice parameters indicates that our Fe$_{3}$O$_{4}$ thin films exhibit a tensile out-of-plane strain of $+1.2$\% and a compressive in-plane strain of $-1.2$\%. This is in contrast to x-ray diffraction measurements published recently.\cite{Vaz:94:2009,Sterbinsky:96:2010} However, those measurements were performed without any electric field applied to the samples,~i.e., in a ferroelastic multi-domain state of the BTO substrate. The derived values of the in-plane and out-of-plane strain yield a Poisson ratio of 0.33, which is in good agreement with Ref.~\onlinecite{Schwenk:13:2000}. In contrast to the x-ray measurements at $E_{z}=400$\,kV/m, the intensity of the Fe$_{3}$O$_{4}$~(004) reflection at $E_{z}=0$\,kV/m is strongly reduced and a second peak occurs close to the first one [Fig.~\ref{fig:Fig5}(c)]. Two different peak functions are needed to reproduce this situation [cf. dashed lines in Fig.~\ref{fig:Fig5}(c)]. While the position of the first one is almost identical to the peak position measured at $E_{z}=400$\,kV/m, the second peak is located at a higher $q_{00L}$-value [cf.~blue arrows in Figs.~\ref{fig:Fig5}(b) and (c)]. This clearly demonstrates that two differently strained regions are present in the Fe$_{3}$O$_{4}$ thin film at $E_{z}=0$\,kV/m. The one region can be attributed to those parts of the Fe$_{3}$O$_{4}$ thin film clamped to ferroelastic $c$ domains of the BTO crystal, while the other to those regions of the Fe$_{3}$O$_{4}$ thin film positioned on top of ferroelastic $a_{2}$ domains. The $a_{2}$ domains induce a tensile strain along the $y$ direction in the overlaying Fe$_{3}$O$_{4}$ thin film, thereby causing a compressive strain along the out-of-plane direction due to elasticity. This shifts the Fe$_{3}$O$_{4}$~(004) reflection to higher $q_{00L}$ values. From the difference in position of the two peaks visible in Fig.~\ref{fig:Fig5}(c), the induced compressive out-of-plane strain of the Fe$_{3}$O$_{4}$ thin film due to the transformation of ferroelastic $c$ domains into $a_{2}$ domains in the BTO crystal can be calculated to $-(0.53\pm0.03)$\%. By employing the Poisson ratio of 0.33, the induced in-plane strains $\epsilon_{xx}^{\mathrm{ind}}=0$\% and $\epsilon_{yy}^{\mathrm{ind}}=(1.07\pm0.06)$\% are derived. These values are in excellent agreement with the theoretical values $\epsilon_{xx}^{\mathrm{BTO}}=0$\% and $\epsilon_{yy}^{\mathrm{BTO}}=1.05$\% caused by domain reorientations from ferroelastic $c$ to $a_{2}$ domains in the BTO crystal. This provides clear evidence for a perfect strain transmission from the BTO substrate into the Fe$_{3}$O$_{4}$ thin film.

Figure~\ref{fig:Fig5} indicates that the large tensile strain is imposed on the Fe$_{3}$O$_{4}$ thin film by the formation of ferroelastic $a_{2}$ domains in the BTO substrate. Due to magnetoelastic effects this strain should strongly affect the magnetization $\mathbf{M}_{a}$ of those parts of the Fe$_{3}$O$_{4}$ film elastically clamped onto these domains. In contrast, since the (004) reflection of the Fe$_{3}$O$_{4}$ thin film positioned on top of ferroelastic $c$ domains stays nearly unaffected as a function of the applied electric field, the magnetization $\mathbf{M}_{c}$ of these regions is expected to depend only very weakly on the applied electric field. Thus, the converse magnetoelectric effects in Fe$_{3}$O$_{4}$/BTO hybrid structures at room temperature can be described in first order by a two-region model
\begin{equation}
   \mathbf{M}(E) = x_{c}(E)\cdot \mathbf{M}_{c}+\left( 1-x_{c}(E)\right)\cdot \mathbf{M}_{a}\, .
\label{eq:ME}
\end{equation}
Here, $x_{c}(E)$ denotes the volume fraction of the ferroelastic $c$ domains in the BTO substrate as a function of the applied electric field, and $(1-x_{c}(E))$ the remaining volume fraction of the emerging ferroelastic $a_{2}$ domains. In this model, the magnetizations $\mathbf{M}_{a}$ and $\mathbf{M}_{c}$ are considered independent of the electric field. Thus, Eq.~\eqref{eq:ME} only considers ferroelastic domain reorientations and does not take into account linear converse piezoelectric effects as well as effects caused by polarization rotations.\cite{Fu:403:2000} This is reasonable, since the inverse linear piezoelectric effect of BTO leads to a maximum in-plane strain, which is three orders of magnitude smaller than the strain caused by ferroelastic domain reorientation. \cite{Gepraegs:96:2010} Furthermore, effects due to ferroelastic domain walls are also not included in the model. To calculate the converse magnetoelectric effects using Eq.~\eqref{eq:ME}, the volume fraction $x_{c}(E)$ as well as the magnetization values $\mathbf{M}_{c}$ and $\mathbf{M}_{a}$ have to be determined.

\section{Volume fraction of ferroelastic domains}
\label{sec:xc}

\begin{figure}[tb]
  \includegraphics[width=\columnwidth]{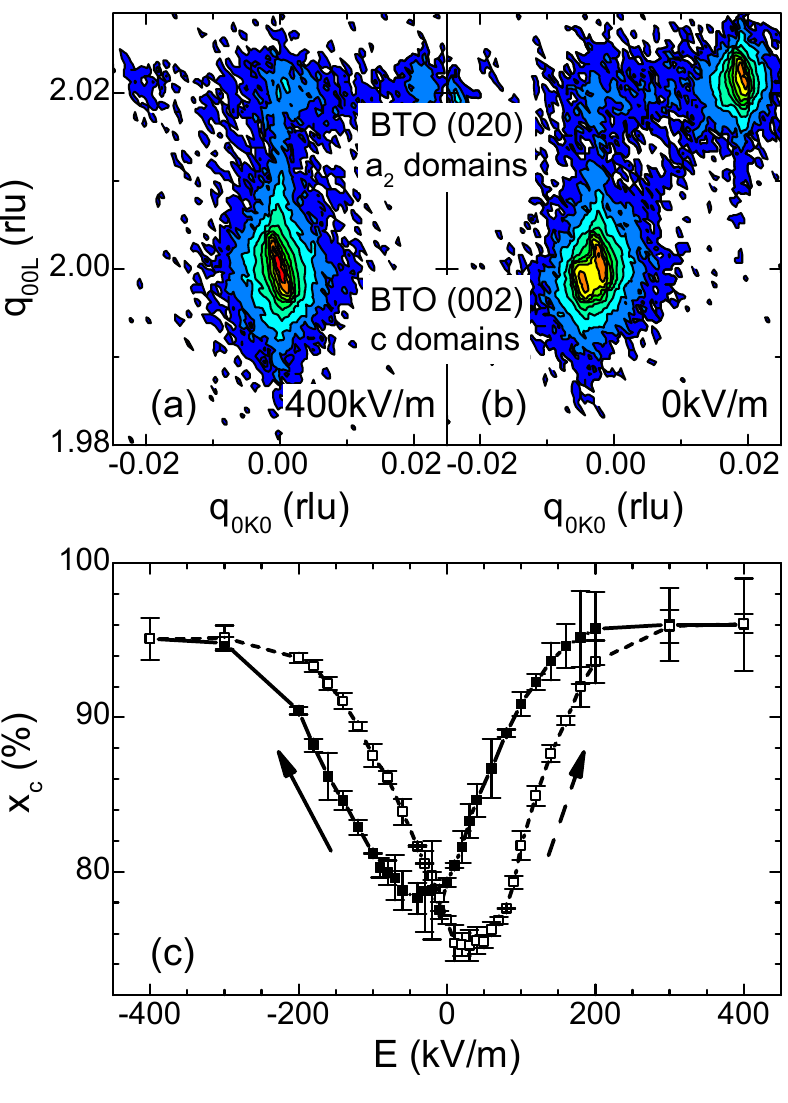}
    \caption{(color online)
Determination of the volume fraction of ferroelastic $c$ domains $x_{c}(E)$ using x-ray diffraction. (a), (b) Mesh scans around the BTO (002) reflection recorded while applying an electric field of 400\,kV/m and 0\,kV/m across the Fe$_{3}$O$_{4}$(38\,nm)/BTO hybrid. To obtain $x_{c}$, the ratio of the integrated intensities for $q_{00L}\leq 2.01$\,rlu and $q_{00L} > 2.01$\,rlu is calculated. (c) $x_{c}$ as a function of the applied electric field. }
    \label{fig:Fig6}
\end{figure}

The volume fraction of the ferroelastic $c$ domains $x_{c}(E)$ is determined by means of x-ray diffraction measurements.\cite{Fong:36:2006} Since the ferroelastic $c$ and $a$ domains exhibit different unit cell dimensions as well as a change of the polarization direction, both the Bragg angles and the intensities of the respective reflections are affected.\cite{Reeuwijk:37:2004} For the experimental determination of $x_c$, x-ray diffraction measurements around the BTO~(002) reflection were carried out for different values of the electric fields applied across the Fe$_{3}$O$_{4}$/BTO hybrid structure.\cite{Subbarao:28:1957} Figures~\ref{fig:Fig6}(a) and (b) show two reciprocal space maps recorded with $E_{z}=400$\,kV/m and $E_{z}=0$\,kV/m. For $E_{z}=400$\,kV/m, only a single reflection around $q_{00L}=2.0$\,rlu is visible, which is a marker for ferroelastic $c$ domains. This demonstrates that at this electric field strength a purely ferroelastic $c$ domain state is present in the BTO substrate. By reducing the electric field strength, tilted $a_{2}$ domains are formed continuously, until the electric coercive field is reached. Scattering from these $a_{2}$ domains causes a finite intensity around $q_{00L}=2.02$\,rlu [cf.~Fig.~\ref{fig:Fig5}(b)]. To determine $x_{c}(E)$, the detected intensities were integrated for $q_{00L}\leq 2.01$\,rlu ($I_{c}$) and $q_{00L}> 2.01$\,rlu ($I_{a}$). Using these intensities, $x_{c}(E)$ can be estimated as $x_{c}(E)=I_{c}/(I_{a}+I_{c})$. The resulting volume fraction as a function of the applied electric field is shown in Fig.~\ref{fig:Fig6}(c). By comparing $x_{c}(E)$ with the $M(E)$-loops depicted in Fig.~\ref{fig:Fig3}, the pronounced interdependence of the magnetization in the ferromagnetic thin film and the volume fraction of the ferroelastic $c$ domains in the BTO substrate becomes obvious. This provides further evidence for the validity of the two-region model described by Eq.~\eqref{eq:ME}. We note, however, that for the calculation of $M(E)$, the ferroelastic domain configuration at the surface of the BTO is essential, since these domains are elastically coupled to the overlaying ferromagnetic thin film. As X-rays penetrate deep into the BTO substrate, the obtained volume fraction $x_{c}(E)$ is an averaged value. By using different angles of the incoming x-ray beam,~i.e.,~varying the penetration depth of the X-rays, an uncertainty of less than 3\% for $x_{c}(E)$ can be estimated. As obvious from Fig.~\ref{fig:Fig6}(c), this uncertainty is not larger than the already displayed error bars, describing the experimental uncertainty at a constant angle of the incoming X-rays.

\section{Magnetoelastic effects}
\label{sec:MaMc}

Having determined the ferroelastic domain state in the BTO substrate as a function of the applied electric field strength, in the two-region model approach [cf.~Eq.~\eqref{eq:ME}] the strain-mediated converse magnetoelectric effects $\mathbf{M}(E)$ in the Fe$_{3}$O$_{4}$/BTO hybrid depend only on the magnetization values $\mathbf{M}_{c}$ and $\mathbf{M}_{a}$ of the regions elastically clamped to ferroelastic $c$ and $a_{2}$ domains, respectively. Therefore, the remaining step is to determine the magnetization values $\mathbf{M}_{c}$ and $\mathbf{M}_{a}$.

In general, elastic strain can modify the magnetization by two main effects. First, the direction of the magnetization can be altered by a strain-induced change of the magnetic anisotropy, since the magnetoelastic energy can be regarded as an additional uniaxial magnetic anisotropy in a phenomenological free energy approach.\cite{Weiler:11:2009} Second, due to strain-induced changes of the bond length and bond angle, the exchange interaction can be modified. This may cause variations of the magnetic ordering temperature and the magnetization. Even a strain-induced transition from a magnetically ordered/electrically conducting to a nonmagnetic/insulating state is possible by changing the spin state of the magnetic ions.\cite{Lengsdorf:69:2004} Since previous experiments showed that changes of the magnetic anisotropy are the driving force behind the converse magnetoelectric effects in Fe$_{3}$O$_{4}$/BTO hybrids,\cite{Gepraegs:86:2012} we neglect any strain-induced modification of the saturation magnetization $M_{\mathrm{s}}$ and assume $M_{\mathrm{s}}$ to be homogeneous throughout the Fe$_{3}$O$_{4}$ thin film. In this approximation, the magnetization values $\mathbf{M}_{c}$ and $\mathbf{M}_{a}$ are described by $\mathbf{M}_{c}=M_{\mathrm{s}}\mathbf{m}_{c}$ and $\mathbf{M}_{a}=M_{\mathrm{s}}\mathbf{m}_{a}$, respectively. Using magnetoelastic theory, the unit vectors $\mathbf{m}_{c}$ and $\mathbf{m}_{a}$ can be derived by minimizing the free enthalpy density $g^{\mathrm{FM}}$ of the Fe$_{3}$O$_{4}$ thin film, which is based on the magnetic energy density $u^{\mathrm{FM}}$ given by (see Ref.~\onlinecite{Gepraegs:86:2012} for more details)
\begin{equation}
   u^{\mathrm{FM}}-u_{0} = u^{\mathrm{FM}}_{\mathrm{ani}}(m_{i})+u^{\mathrm{FM}}_{\mathrm{el}}(\eta_{k}) + u^{\mathrm{FM}}_{\mathrm{magel}}(m_{i},\eta_{k}) + \ldots 
\label{eq:free_energy}
\end{equation}
Here, $u^{\mathrm{FM}}_{\mathrm{ani}}(m_{i})$ is the magnetic anisotropy contribution, $u_{\mathrm{el}}^{\mathrm{FM}}(\eta_{k})$ the purely elastic energy density, $u^{\mathrm{FM}}_{\mathrm{magel}}$ the magnetoelastic energy density, and $m_{i}$ with $i=1,2,3$ are the components of the unit vectors $\mathbf{m}_{a,c}$,~i.e., the directional cosines. The strain components $\eta_{k}$ ($k=1,\ldots ,6$) are given in matrix notation with $\eta_{1}=\epsilon_{11}$, $\eta_{2}=\epsilon_{22}$, $\eta_{3}=\epsilon_{33}$, $\eta_{4}=2\epsilon_{23}$, $\eta_{5}=2\epsilon_{31}$, and $\eta_{6}=2\epsilon_{12}$.\cite{Nye:book} To account for the shape anisotropy in ferromagnetic thin films with finite dimensions, an additional contribution $u^{\mathrm{FM}}_{\mathrm{demag}}(m_{3})=(\mu_{0}/2)M^2_{\mathrm{s}}m^2_{3}$ is added to Eq.~\eqref{eq:free_energy}.

In the framework of a magnetic single domain model, $u^{\mathrm{FM}}$ is only a function of the magnetization direction $m_{i}$ and the strain components $\eta_{k}$. The components $\eta_{k}$ with $k\geq4$ are zero, since no shear strains are expected in the tetragonal phase of BTO at room temperature. As discussed in Section~\ref{sec:model}, the strain state of those parts of the Fe$_{3}$O$_{4}$ thin film elastically coupled to ferroelastic $c$ domains ($\eta_{c}$) and $a$ domains ($\eta_{a}$) of the BTO substrate is
\begin{equation}
\eta_{c}=
\left(\begin{array}{c}
-0.012\\
-0.012\\
+0.012\\
0\\
0\\
0
\end{array}\right), \;\;\;\; 
\eta_{a}=
\left(\begin{array}{c}
-0.012\\
-0.001\\
+0.005\\
0\\
0\\
0
\end{array}\right).
\end{equation}
Assuming a cubic symmetry of the Fe$_{3}$O$_{4}$ thin film, the magnetic anisotropy $u^{\mathrm{FM}}_{\mathrm{ani}}(m_{i})$ can be expressed as $u^{\mathrm{FM}}_{\mathrm{ani}}(m_{i})=K_{c}\left(m_{1}^{2}m_{2}^{2}+m_{2}^{2}m_{3}^{2}+m_{3}^{2}m_{1}^{2}\right)$, with the first order cubic anisotropy constant $K_{c}$. The second term in Eq.~\eqref{eq:free_energy} represents the purely elastic energy $u_{\mathrm{el}}^{\mathrm{FM}}(\eta_{k})=\frac{1}{2}c_{11}\left(\eta^{2}_{1} + \eta^{2}_{2}+\eta^{2}_{3}\right)+c_{12}\left(\eta_{1}\eta_{2}+\eta_{2}\eta_{3}+\eta_{1}\eta_{3}\right) + \frac{1}{2}c_{44}\left(\eta_{4}^{2}+\eta_{5}^{2}+\eta_{6}^{2}\right)$. Here, $c_{ij}$ are the elastic stiffness constants of the cubic Fe$_{3}$O$_{4}$ thin film, which are regarded as material constants ($c_{11}=27.2 \times 10^{10}$\,N/m$^{2}$, $c_{12}=17.8 \times 10^{10}$\,N/m$^{2}$, and $c_{44}=6.1 \times 10^{10}$\,N/cm$^{2}$).\cite{Schwenk:13:2000} The interaction between the elastic and the magnetic anisotropy energies is described by the first order magnetoelastic energy density $u^{\mathrm{FM}}_{\mathrm{magel}}$:
\begin{eqnarray}
u^{\mathrm{FM}}_{\mathrm{magel}} & = & \chi \left[ \vphantom{\sum_{j}\eta_{i}} B_{1}\sum_{i}\eta_{i}\left( m_{i}^{2}-1/3 \right) \right. 
\nonumber \\
 & & + \left. B_{2} \sum_{i>j} \eta_{9-i+j} m_{i}m_{j} \right]\, ,
    \label{eq:magel}
\end{eqnarray}
with the magnetoelastic coupling coefficients $B_{1}$ and $B_{2}$. For bulk magnetostrictive samples, $B_{1}$ and $B_{2}$ can be expressed as a function of the magnetostrictive strains $\lambda_{100}$ and $\lambda_{111}$: $B_{1}=-\frac{3}{2}\lambda_{100}\left(c_{11}-c_{12}\right)$ and $B_{2}=-3\lambda_{111}c_{44}$. We note that the magnetoelastic coupling coefficients $B_{i}$ in ferromagnetic thin films may deviate from the bulk values due to surface effects and/or the influence of strain.\cite{Tian:79:2009,Sun:66:1991} In our analysis, we neglect these effects and use the bulk values $\lambda_{100}=-19.5 \times 10^{-6}$ and $\lambda_{111}=+77.6 \times 10^{-6}$ for the calculations.\cite{Bickford:99:1210} A proportionality factor $\chi$ is introduced to account for any deviation in the magnetoelastic coupling from bulk-like behavior.

\begin{figure}[tb]
  \includegraphics[width=\columnwidth]{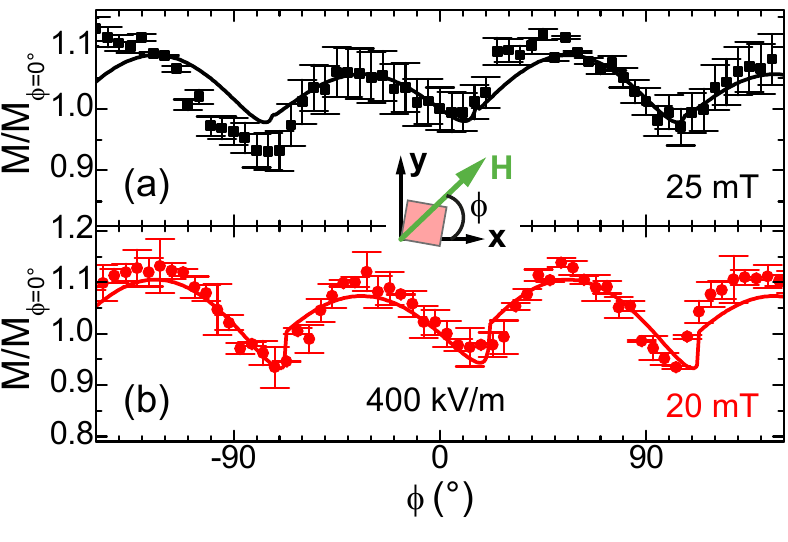}
    \caption{(color online)
Magnetization $M$ plotted versus the angle $\phi$ defining the in-plane orientation of the applied magnetic field. $\phi=0^\circ$ corresponds to the $x$ direction. The data are taken at an electric field of $E_{z}=400$\,kV/m and an external magnetic field of 25\,mT (black symbols) and 20\,mT (red symbols). The $M(\phi)$-curves can be fitted using a cubic anisotropy field of $K_{c}/M_{\mathrm{s}}=(-19 \pm 1)$\,mT (solid lines), assuming a misalignment of $4^\circ$ between the sample normal and the rotation axis.}
    \label{fig:Fig7}
\end{figure}

To determine the cubic anisotropy field $K_{c}/M_{\mathrm{s}}$, SQUID magnetometry measurements were carried out for different in-plane orientations of the applied magnetic field. To ensure a well defined ferroelastic single $c$ domain state in the BTO substrate, an electric field of $E_{z}=400$\,kV/m was applied across the hybrid structure. The angular dependence of the magnetization projection $M(\phi)/M(\phi=0^\circ)$ along the external magnetic field $H$ is shown in Fig.~\ref{fig:Fig7} for $\mu_{0}H=20$\,mT and $\mu_{0}H=25$\,mT. Obviously, a four-fold symmetry is visible, which is a clear sign of a cubic magnetic anisotropy in the Fe$_{3}$O$_{4}$ thin film. The angular dependence $M(\phi)/M(\phi=0^\circ)$ measured at constant strain state $\eta_{c}$ can be simulated by means of Eq.~\eqref{eq:free_energy}. The best fit between experiment and simulation was obtained using a cubic anisotropy field of $K_{c}/M_{\mathrm{s}}=(-19 \pm 1)$\,mT (cf.~solid lines in Fig.~\ref{fig:Fig7}). A misalignment of $4^\circ$ between the sample normal and the rotation axis was assumed to account for the different magnetization values measured at $\phi \approx -80^\circ$, $\phi \approx 10^\circ$, and $\phi \approx 100^\circ$ (cf.~Fig.~\ref{fig:Fig7}).

Using the cubic anisotropy field and the elastic parameters determined above, the Fe$_{3}$O$_{4}$/BTO hybrid structure is well described except for the remaining unknown parameter $\chi$, which characterizes the magnetoelastic coupling strength. This parameter is the only fitting parameter for the simulation of the converse magnetoelectric effect in the following.

\section{Experiment versus simulation}
\label{sec:ExpSimu}

\begin{figure}[tb]
  \includegraphics[width=\columnwidth]{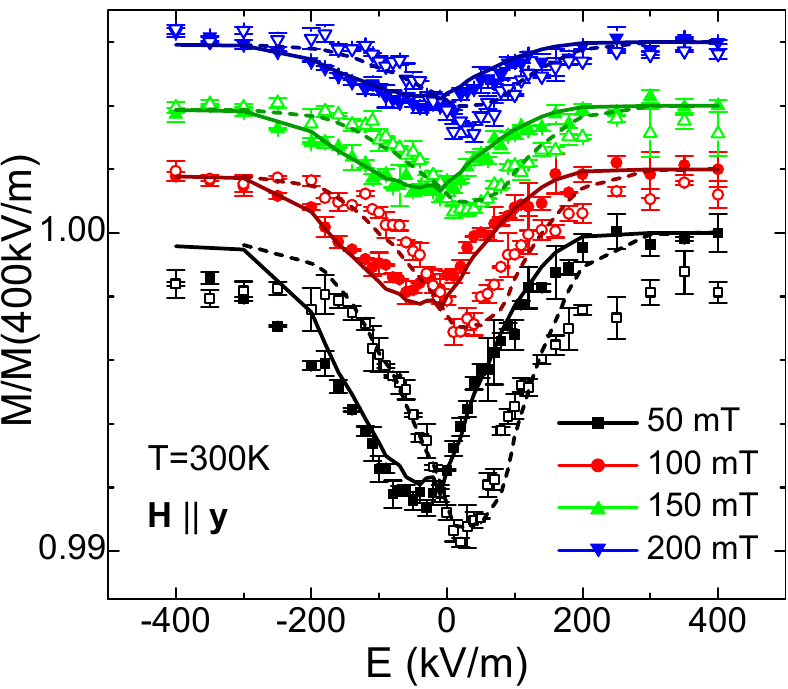}
    \caption{(color online)
Measurement (symbols) and simulation (lines) of the converse magnetoelectric effect in a Fe$_{3}$O$_{4}$(38\,nm)/BTO multiferroic hybrid for different magnetic field strengths. The magnetic field was aligned in the film plane along the $y$ direction of the BTO crystal and the temperature was set to 300\,K. For better visibility, the curves are vertically shifted with respect to each other. The best fit between experiment and simulation was obtained by assuming a magnetoelastic coupling efficiency of 55\% ($\chi=0.55$). A misalignment of the sample surface with respect to the magnetic field of $5^\circ$ was assumed for the calculation.}
    \label{fig:Fig8}
\end{figure}

By calculating the magnetization values $\mathbf{M}_{c}$ and $\mathbf{M}_{a}$ according to  Eq.~\eqref{eq:free_energy} the converse magnetoelectric effects in the Fe$_{3}$O$_{4}$(38\,nm)/BTO hybrid (cf.~Fig.~\ref{fig:Fig3}) can be simulated for different magnetic field strengths assuming that ferroelastic domain reorientation is the dominant contribution [cf.~Eq.~\eqref{eq:ME}]. The normalized magnetization projection along the $y$ direction as a function of the applied electric field obtained in this way is shown in Fig.~\ref{fig:Fig8}, together with the experimental results taken from Fig.~\ref{fig:Fig3}. An excellent agreement between experiment and simulation could be obtained by assuming a magnetoelastic coupling efficiency of 55\% ($\chi=0.55$). The fact that this value is well below 100\% indicates that the magnetoelastic behavior of the Fe$_{3}$O$_{4}$ thin film is quite different from the bulk. Nevertheless, as obvious from Fig.~\ref{fig:Fig8}, the variation of the magnetization with the applied electric field can be well simulated for magnetic fields $\mu_{0}H\geq50$\,mT using the two-region model described in Eq.~\eqref{eq:ME}. At magnetic field values close to the coercive field, converse piezoelectric effects and magnetic domain effects come into play, leading to large changes of the magnetic coercivity as a function of the electric field.\cite{Sahoo:76:2007,Gepraegs:96:2010} These changes can not be explained within the simple two-region model alone. In this case, the two-region model must be further extended to take into account piezoelectric effects as well as strain effects due to polarization rotations to simulate the electrical manipulation of the magnetic coercive fields.

\section{Conclusion}

We have investigated strain-mediated converse magnetoelectric effects in extrinsic multiferroic hybrid structures consisting of epitaxially grown ferrimagnetic Fe$_{3}$O$_{4}$ thin films on ferroelectric BTO substrates at room temperature. As evident from the time evolution of the intensity recorded by in-situ RHEED, the growth of Fe$_{3}$O$_{4}$ thin films onto BTO substrates is highly disordered for the first few monolayers and then continuously evolves into a true layer-by-layer growth mode. SQUID magnetometry measurements on these Fe$_{3}$O$_{4}$ thin films reveal a saturation magnetization close to the bulk value as well as a pronounced Verwey transition, which is a hallmark of the quality of Fe$_{3}$O$_{4}$. Furthermore, butterfly-shaped changes of the magnetization as a function of the applied electric field were observed at room temperature. The use of miscut BTO substrates, where only a single type of ferroelastic $a$ domains exist, allows us to quantitatively simulate the observed converse magnetoelectric effect by using a simple two-region model. This model considers only effects caused by reorientations of ferroelastic $c$ domains into ferroelastic $a$ domains in the BTO substrate. In the framework of this model, the magnetization values of those parts of the Fe$_{3}$O$_{4}$ thin film elastically clamped onto ferroelastic $c$ domains and $a$ domains are considered as electric field independent quantities, which can be calculated using magnetoelastic theory. By comparing the results of the simulations to the experimental data, we demonstrate that the room-temperature converse magnetoelectric effects of Fe$_{3}$O$_{4}$/BTO-based multiferroic hybrid structures can be well understood on the basis of the two-region model for applied magnetic fields larger than the saturation field of Fe$_{3}$O$_{4}$. The detailed understanding of the converse magnetoelectric effect at room temperature in BTO-based multiferroic hybrid structures opens the way to develop novel extrinsic multiferroic hybrids exhibiting giant electric-field induced changes of the magnetization.

\section*{Acknowledgment}

We thank Thomas Brenninger for continuous technical support. Financial support from the German Research Foundation (projects GR 1132/13 and 14) and the German Excellence Initiative via the ''Nanosystems Initiative Munich'' (NIM) is gratefully acknowledged.


\begin{thebibliography}{87}%
\makeatletter
\providecommand \@ifxundefined [1]{%
 \@ifx{#1\undefined}
}%
\providecommand \@ifnum [1]{%
 \ifnum #1\expandafter \@firstoftwo
 \else \expandafter \@secondoftwo
 \fi
}%
\providecommand \@ifx [1]{%
 \ifx #1\expandafter \@firstoftwo
 \else \expandafter \@secondoftwo
 \fi
}%
\providecommand \natexlab [1]{#1}%
\providecommand \enquote  [1]{``#1''}%
\providecommand \bibnamefont  [1]{#1}%
\providecommand \bibfnamefont [1]{#1}%
\providecommand \citenamefont [1]{#1}%
\providecommand \href@noop [0]{\@secondoftwo}%
\providecommand \href [0]{\begingroup \@sanitize@url \@href}%
\providecommand \@href[1]{\@@startlink{#1}\@@href}%
\providecommand \@@href[1]{\endgroup#1\@@endlink}%
\providecommand \@sanitize@url [0]{\catcode `\\12\catcode `\$12\catcode
  `\&12\catcode `\#12\catcode `\^12\catcode `\_12\catcode `\%12\relax}%
\providecommand \@@startlink[1]{}%
\providecommand \@@endlink[0]{}%
\providecommand \url  [0]{\begingroup\@sanitize@url \@url }%
\providecommand \@url [1]{\endgroup\@href {#1}{\urlprefix }}%
\providecommand \urlprefix  [0]{URL }%
\providecommand \Eprint [0]{\href }%
\providecommand \doibase [0]{http://dx.doi.org/}%
\providecommand \selectlanguage [0]{\@gobble}%
\providecommand \bibinfo  [0]{\@secondoftwo}%
\providecommand \bibfield  [0]{\@secondoftwo}%
\providecommand \translation [1]{[#1]}%
\providecommand \BibitemOpen [0]{}%
\providecommand \bibitemStop [0]{}%
\providecommand \bibitemNoStop [0]{.\EOS\space}%
\providecommand \EOS [0]{\spacefactor3000\relax}%
\providecommand \BibitemShut  [1]{\csname bibitem#1\endcsname}%
\let\auto@bib@innerbib\@empty
\bibitem [{\citenamefont {Binek}\ and\ \citenamefont
  {Doudin}(2005)}]{Binek:17:2005}%
  \BibitemOpen
  \bibfield  {author} {\bibinfo {author} {\bibfnamefont {C.}~\bibnamefont
  {Binek}}\ and\ \bibinfo {author} {\bibfnamefont {B.}~\bibnamefont {Doudin}},\
  }\href {\doibase 10.1088/0953-8984/17/2/L06} {\bibfield  {journal} {\bibinfo
  {journal} {J. Phys.: Condens. Matter}\ }\textbf {\bibinfo {volume} {17}},\
  \bibinfo {pages} {L39} (\bibinfo {year} {2005})}\BibitemShut {NoStop}%
\bibitem [{\citenamefont {Eerenstein}\ \emph {et~al.}(2006)\citenamefont
  {Eerenstein}, \citenamefont {Mathur},\ and\ \citenamefont
  {Scott}}]{Eerenstein:442:2006}%
  \BibitemOpen
  \bibfield  {author} {\bibinfo {author} {\bibfnamefont {W.}~\bibnamefont
  {Eerenstein}}, \bibinfo {author} {\bibfnamefont {N.~D.}\ \bibnamefont
  {Mathur}}, \ and\ \bibinfo {author} {\bibfnamefont {J.~F.}\ \bibnamefont
  {Scott}},\ }\href {\doibase 10.1038/nature05023} {\bibfield  {journal}
  {\bibinfo  {journal} {Nature}\ }\textbf {\bibinfo {volume} {442}},\ \bibinfo
  {pages} {759} (\bibinfo {year} {2006})}\BibitemShut {NoStop}%
\bibitem [{\citenamefont {Ramesh}\ and\ \citenamefont
  {Spaldin}(2007)}]{Ramesh:6:2007}%
  \BibitemOpen
  \bibfield  {author} {\bibinfo {author} {\bibfnamefont {R.}~\bibnamefont
  {Ramesh}}\ and\ \bibinfo {author} {\bibfnamefont {N.~A.}\ \bibnamefont
  {Spaldin}},\ }\href {\doibase 10.1038/nmat1805} {\bibfield  {journal}
  {\bibinfo  {journal} {Nat. Mater.}\ }\textbf {\bibinfo {volume} {6}},\
  \bibinfo {pages} {21} (\bibinfo {year} {2007})}\BibitemShut {NoStop}%
\bibitem [{\citenamefont {Goennenwein}\ \emph {et~al.}(2008)\citenamefont
  {Goennenwein}, \citenamefont {Althammer}, \citenamefont {Bihler},
  \citenamefont {Brandlmaier}, \citenamefont {Gepr{\"a}gs}, \citenamefont
  {Opel}, \citenamefont {Schoch}, \citenamefont {Limmer}, \citenamefont
  {Gross},\ and\ \citenamefont {Brandt}}]{Goennenwein:2:2008}%
  \BibitemOpen
  \bibfield  {author} {\bibinfo {author} {\bibfnamefont {S.~T.~B.}\
  \bibnamefont {Goennenwein}}, \bibinfo {author} {\bibfnamefont
  {M.}~\bibnamefont {Althammer}}, \bibinfo {author} {\bibfnamefont
  {C.}~\bibnamefont {Bihler}}, \bibinfo {author} {\bibfnamefont
  {A.}~\bibnamefont {Brandlmaier}}, \bibinfo {author} {\bibfnamefont
  {S.}~\bibnamefont {Gepr{\"a}gs}}, \bibinfo {author} {\bibfnamefont
  {M.}~\bibnamefont {Opel}}, \bibinfo {author} {\bibfnamefont {W.}~\bibnamefont
  {Schoch}}, \bibinfo {author} {\bibfnamefont {W.}~\bibnamefont {Limmer}},
  \bibinfo {author} {\bibfnamefont {R.}~\bibnamefont {Gross}}, \ and\ \bibinfo
  {author} {\bibfnamefont {M.~S.}\ \bibnamefont {Brandt}},\ }\href {\doibase
  10.1002/pssr.200802011} {\bibfield  {journal} {\bibinfo  {journal} {Phys.
  Stat. Sol. (RRL)}\ }\textbf {\bibinfo {volume} {2}},\ \bibinfo {pages} {96}
  (\bibinfo {year} {2008})}\BibitemShut {NoStop}%
\bibitem [{\citenamefont {Weiler}\ \emph {et~al.}(2009)\citenamefont {Weiler},
  \citenamefont {Brandlmaier}, \citenamefont {Gepr\"ags}, \citenamefont
  {Althammer}, \citenamefont {Opel}, \citenamefont {Bihler}, \citenamefont
  {Huebl}, \citenamefont {Brandt}, \citenamefont {Gross},\ and\ \citenamefont
  {Goennenwein}}]{Weiler:11:2009}%
  \BibitemOpen
  \bibfield  {author} {\bibinfo {author} {\bibfnamefont {M.}~\bibnamefont
  {Weiler}}, \bibinfo {author} {\bibfnamefont {A.}~\bibnamefont {Brandlmaier}},
  \bibinfo {author} {\bibfnamefont {S.}~\bibnamefont {Gepr\"ags}}, \bibinfo
  {author} {\bibfnamefont {M.}~\bibnamefont {Althammer}}, \bibinfo {author}
  {\bibfnamefont {M.}~\bibnamefont {Opel}}, \bibinfo {author} {\bibfnamefont
  {C.}~\bibnamefont {Bihler}}, \bibinfo {author} {\bibfnamefont
  {H.}~\bibnamefont {Huebl}}, \bibinfo {author} {\bibfnamefont {M.~S.}\
  \bibnamefont {Brandt}}, \bibinfo {author} {\bibfnamefont {R.}~\bibnamefont
  {Gross}}, \ and\ \bibinfo {author} {\bibfnamefont {S.~T.~B.}\ \bibnamefont
  {Goennenwein}},\ }\href {\doibase 10.1088/1367-2630/11/1/013021} {\bibfield
  {journal} {\bibinfo  {journal} {New J. Phys.}\ }\textbf {\bibinfo {volume}
  {11}},\ \bibinfo {pages} {013021} (\bibinfo {year} {2009})}\BibitemShut
  {NoStop}%
\bibitem [{\citenamefont {Brandlmaier}\ \emph {et~al.}(2011)\citenamefont
  {Brandlmaier}, \citenamefont {Gepr\"{a}gs}, \citenamefont {Woltersdorf},
  \citenamefont {Gross},\ and\ \citenamefont
  {Goennenwein}}]{Brandlmaier:110:2011}%
  \BibitemOpen
  \bibfield  {author} {\bibinfo {author} {\bibfnamefont {A.}~\bibnamefont
  {Brandlmaier}}, \bibinfo {author} {\bibfnamefont {S.}~\bibnamefont
  {Gepr\"{a}gs}}, \bibinfo {author} {\bibfnamefont {G.}~\bibnamefont
  {Woltersdorf}}, \bibinfo {author} {\bibfnamefont {R.}~\bibnamefont {Gross}},
  \ and\ \bibinfo {author} {\bibfnamefont {S.~T.~B.}\ \bibnamefont
  {Goennenwein}},\ }\href {\doibase 10.1063/1.3624663} {\bibfield  {journal}
  {\bibinfo  {journal} {J. Appl. Phys.}\ }\textbf {\bibinfo {volume} {110}},\
  \bibinfo {pages} {043913} (\bibinfo {year} {2011})}\BibitemShut {NoStop}%
\bibitem [{\citenamefont {Hu}\ \emph {et~al.}(2011)\citenamefont {Hu},
  \citenamefont {Li}, \citenamefont {Chen},\ and\ \citenamefont
  {Nan}}]{Hu:2:2011}%
  \BibitemOpen
  \bibfield  {author} {\bibinfo {author} {\bibfnamefont {J.-M.}\ \bibnamefont
  {Hu}}, \bibinfo {author} {\bibfnamefont {Z.}~\bibnamefont {Li}}, \bibinfo
  {author} {\bibfnamefont {L.-Q.}\ \bibnamefont {Chen}}, \ and\ \bibinfo
  {author} {\bibfnamefont {C.-W.}\ \bibnamefont {Nan}},\ }\href {\doibase
  10.1038/ncomms1564} {\bibfield  {journal} {\bibinfo  {journal} {Nature
  Communications}\ }\textbf {\bibinfo {volume} {2}},\ \bibinfo {pages} {553}
  (\bibinfo {year} {2011})}\BibitemShut {NoStop}%
\bibitem [{\citenamefont {Heron}\ \emph {et~al.}(2011)\citenamefont {Heron},
  \citenamefont {Trassin}, \citenamefont {Ashraf}, \citenamefont {Gajek},
  \citenamefont {He}, \citenamefont {Yang}, \citenamefont {Nikonov},
  \citenamefont {Chu}, \citenamefont {Salahuddin},\ and\ \citenamefont
  {Ramesh}}]{Heron:107:2011}%
  \BibitemOpen
  \bibfield  {author} {\bibinfo {author} {\bibfnamefont {J.~T.}\ \bibnamefont
  {Heron}}, \bibinfo {author} {\bibfnamefont {M.}~\bibnamefont {Trassin}},
  \bibinfo {author} {\bibfnamefont {K.}~\bibnamefont {Ashraf}}, \bibinfo
  {author} {\bibfnamefont {M.}~\bibnamefont {Gajek}}, \bibinfo {author}
  {\bibfnamefont {Q.}~\bibnamefont {He}}, \bibinfo {author} {\bibfnamefont
  {S.~Y.}\ \bibnamefont {Yang}}, \bibinfo {author} {\bibfnamefont {D.~E.}\
  \bibnamefont {Nikonov}}, \bibinfo {author} {\bibfnamefont {Y.-H.}\
  \bibnamefont {Chu}}, \bibinfo {author} {\bibfnamefont {S.}~\bibnamefont
  {Salahuddin}}, \ and\ \bibinfo {author} {\bibfnamefont {R.}~\bibnamefont
  {Ramesh}},\ }\href {\doibase 10.1103/PhysRevLett.107.217202} {\bibfield
  {journal} {\bibinfo  {journal} {Phys. Rev. Lett.}\ }\textbf {\bibinfo
  {volume} {107}},\ \bibinfo {pages} {217202} (\bibinfo {year}
  {2011})}\BibitemShut {NoStop}%
\bibitem [{\citenamefont {Hu}\ \emph {et~al.}(2012)\citenamefont {Hu},
  \citenamefont {Li}, \citenamefont {Chen},\ and\ \citenamefont
  {Nan}}]{Hu:24:2012}%
  \BibitemOpen
  \bibfield  {author} {\bibinfo {author} {\bibfnamefont {J.-M.}\ \bibnamefont
  {Hu}}, \bibinfo {author} {\bibfnamefont {Z.}~\bibnamefont {Li}}, \bibinfo
  {author} {\bibfnamefont {L.-Q.}\ \bibnamefont {Chen}}, \ and\ \bibinfo
  {author} {\bibfnamefont {C.-W.}\ \bibnamefont {Nan}},\ }\href {\doibase
  10.1002/adma.201201004} {\bibfield  {journal} {\bibinfo  {journal} {Adv.
  Mater.}\ }\textbf {\bibinfo {volume} {24}},\ \bibinfo {pages} {2869}
  (\bibinfo {year} {2012})}\BibitemShut {NoStop}%
\bibitem [{\citenamefont {Vaz}(2012)}]{Vaz:24:2012}%
  \BibitemOpen
  \bibfield  {author} {\bibinfo {author} {\bibfnamefont {C.~A.~F.}\
  \bibnamefont {Vaz}},\ }\href {\doibase 10.1088/0953-8984/24/33/333201}
  {\bibfield  {journal} {\bibinfo  {journal} {J. Phys.: Condens. Matter}\
  }\textbf {\bibinfo {volume} {24}},\ \bibinfo {pages} {333201} (\bibinfo
  {year} {2012})}\BibitemShut {NoStop}%
\bibitem [{\citenamefont {Gepr{\"a}gs}\ \emph {et~al.}(2007)\citenamefont
  {Gepr{\"a}gs}, \citenamefont {Opel}, \citenamefont {Goennenwein},\ and\
  \citenamefont {Gross}}]{Gepraegs:87:2007}%
  \BibitemOpen
  \bibfield  {author} {\bibinfo {author} {\bibfnamefont {S.}~\bibnamefont
  {Gepr{\"a}gs}}, \bibinfo {author} {\bibfnamefont {M.}~\bibnamefont {Opel}},
  \bibinfo {author} {\bibfnamefont {S.~T.~B.}\ \bibnamefont {Goennenwein}}, \
  and\ \bibinfo {author} {\bibfnamefont {R.}~\bibnamefont {Gross}},\ }\href
  {\doibase 10.1080/09500830701194165} {\bibfield  {journal} {\bibinfo
  {journal} {Philos. Mag. Lett.}\ }\textbf {\bibinfo {volume} {87}},\ \bibinfo
  {pages} {141} (\bibinfo {year} {2007})}\BibitemShut {NoStop}%
\bibitem [{\citenamefont {Nan}\ \emph {et~al.}(2008)\citenamefont {Nan},
  \citenamefont {Bichurin}, \citenamefont {Dong}, \citenamefont {Viehland},\
  and\ \citenamefont {Srinivasan}}]{Nan:103:2008}%
  \BibitemOpen
  \bibfield  {author} {\bibinfo {author} {\bibfnamefont {C.-W.}\ \bibnamefont
  {Nan}}, \bibinfo {author} {\bibfnamefont {M.~I.}\ \bibnamefont {Bichurin}},
  \bibinfo {author} {\bibfnamefont {S.}~\bibnamefont {Dong}}, \bibinfo {author}
  {\bibfnamefont {D.}~\bibnamefont {Viehland}}, \ and\ \bibinfo {author}
  {\bibfnamefont {G.}~\bibnamefont {Srinivasan}},\ }\href {\doibase
  10.1063/1.2836410} {\bibfield  {journal} {\bibinfo  {journal} {J. Appl.
  Phys.}\ }\textbf {\bibinfo {volume} {103}},\ \bibinfo {pages} {031101}
  (\bibinfo {year} {2008})}\BibitemShut {NoStop}%
\bibitem [{\citenamefont {Vaz}\ \emph {et~al.}(2010)\citenamefont {Vaz},
  \citenamefont {Hoffman}, \citenamefont {Ahn},\ and\ \citenamefont
  {Ramesh}}]{Vaz:22:2010}%
  \BibitemOpen
  \bibfield  {author} {\bibinfo {author} {\bibfnamefont {C.~A.~F.}\
  \bibnamefont {Vaz}}, \bibinfo {author} {\bibfnamefont {J.}~\bibnamefont
  {Hoffman}}, \bibinfo {author} {\bibfnamefont {C.~H.}\ \bibnamefont {Ahn}}, \
  and\ \bibinfo {author} {\bibfnamefont {R.}~\bibnamefont {Ramesh}},\ }\href
  {\doibase 10.1002/adma.200904326} {\bibfield  {journal} {\bibinfo  {journal}
  {Adv. Mater.}\ }\textbf {\bibinfo {volume} {22}},\ \bibinfo {pages} {2900}
  (\bibinfo {year} {2010})}\BibitemShut {NoStop}%
\bibitem [{\citenamefont {Ma}\ \emph {et~al.}(2011)\citenamefont {Ma},
  \citenamefont {Hu}, \citenamefont {Li},\ and\ \citenamefont
  {Nan}}]{Ma:AdvMater:23:2011}%
  \BibitemOpen
  \bibfield  {author} {\bibinfo {author} {\bibfnamefont {J.}~\bibnamefont
  {Ma}}, \bibinfo {author} {\bibfnamefont {J.}~\bibnamefont {Hu}}, \bibinfo
  {author} {\bibfnamefont {Z.}~\bibnamefont {Li}}, \ and\ \bibinfo {author}
  {\bibfnamefont {C.-W.}\ \bibnamefont {Nan}},\ }\href {\doibase
  10.1002/adma.201190024} {\bibfield  {journal} {\bibinfo  {journal} {Adv.
  Mater.}\ }\textbf {\bibinfo {volume} {23}},\ \bibinfo {pages} {1061}
  (\bibinfo {year} {2011})}\BibitemShut {NoStop}%
\bibitem [{\citenamefont {Brandlmaier}\ \emph {et~al.}(2012)\citenamefont
  {Brandlmaier}, \citenamefont {Brasse}, \citenamefont {Gepr\"ags},
  \citenamefont {Weiler}, \citenamefont {Gross},\ and\ \citenamefont
  {Goennenwein}}]{Brandlmaier:85:2012}%
  \BibitemOpen
  \bibfield  {author} {\bibinfo {author} {\bibfnamefont {A.}~\bibnamefont
  {Brandlmaier}}, \bibinfo {author} {\bibfnamefont {M.}~\bibnamefont {Brasse}},
  \bibinfo {author} {\bibfnamefont {S.}~\bibnamefont {Gepr\"ags}}, \bibinfo
  {author} {\bibfnamefont {M.}~\bibnamefont {Weiler}}, \bibinfo {author}
  {\bibfnamefont {R.}~\bibnamefont {Gross}}, \ and\ \bibinfo {author}
  {\bibfnamefont {S.~T.~B.}\ \bibnamefont {Goennenwein}},\ }\href {\doibase
  10.1140/epjb/e2012-20675-4} {\bibfield  {journal} {\bibinfo  {journal} {Eur.
  Phys. J. B}\ }\textbf {\bibinfo {volume} {85}} (\bibinfo {year} {2012}),\
  10.1140/epjb/e2012-20675-4}\BibitemShut {NoStop}%
\bibitem [{\citenamefont {Chiba}\ \emph {et~al.}(2011)\citenamefont {Chiba},
  \citenamefont {Fukami}, \citenamefont {Shimamura}, \citenamefont {Ishiwata},
  \citenamefont {Kobayashi},\ and\ \citenamefont {Ono}}]{Chiba:10:2011}%
  \BibitemOpen
  \bibfield  {author} {\bibinfo {author} {\bibfnamefont {D.}~\bibnamefont
  {Chiba}}, \bibinfo {author} {\bibfnamefont {S.}~\bibnamefont {Fukami}},
  \bibinfo {author} {\bibfnamefont {K.}~\bibnamefont {Shimamura}}, \bibinfo
  {author} {\bibfnamefont {N.}~\bibnamefont {Ishiwata}}, \bibinfo {author}
  {\bibfnamefont {K.}~\bibnamefont {Kobayashi}}, \ and\ \bibinfo {author}
  {\bibfnamefont {T.}~\bibnamefont {Ono}},\ }\href {\doibase 10.1038/nmat3130}
  {\bibfield  {journal} {\bibinfo  {journal} {Nat. Mater.}\ }\textbf {\bibinfo
  {volume} {10}},\ \bibinfo {pages} {853} (\bibinfo {year} {2011})}\BibitemShut
  {NoStop}%
\bibitem [{\citenamefont {Molegraaf}\ \emph {et~al.}(2009)\citenamefont
  {Molegraaf}, \citenamefont {Hoffman}, \citenamefont {Vaz}, \citenamefont
  {Gariglio}, \citenamefont {van~der Marel}, \citenamefont {Ahn},\ and\
  \citenamefont {Triscone}}]{Molegraaf:21:2009}%
  \BibitemOpen
  \bibfield  {author} {\bibinfo {author} {\bibfnamefont {H.~J.~A.}\
  \bibnamefont {Molegraaf}}, \bibinfo {author} {\bibfnamefont {J.}~\bibnamefont
  {Hoffman}}, \bibinfo {author} {\bibfnamefont {C.~A.~F.}\ \bibnamefont {Vaz}},
  \bibinfo {author} {\bibfnamefont {S.}~\bibnamefont {Gariglio}}, \bibinfo
  {author} {\bibfnamefont {D.}~\bibnamefont {van~der Marel}}, \bibinfo {author}
  {\bibfnamefont {C.~H.}\ \bibnamefont {Ahn}}, \ and\ \bibinfo {author}
  {\bibfnamefont {J.-M.}\ \bibnamefont {Triscone}},\ }\href {\doibase
  10.1002/adma.200900278} {\bibfield  {journal} {\bibinfo  {journal} {Adv.
  Mater.}\ }\textbf {\bibinfo {volume} {21}},\ \bibinfo {pages} {3470}
  (\bibinfo {year} {2009})}\BibitemShut {NoStop}%
\bibitem [{\citenamefont {Hochstrat}\ \emph {et~al.}(2004)\citenamefont
  {Hochstrat}, \citenamefont {Binek}, \citenamefont {Chen},\ and\ \citenamefont
  {Kleemann}}]{Hochstrat:272:2004}%
  \BibitemOpen
  \bibfield  {author} {\bibinfo {author} {\bibfnamefont {A.}~\bibnamefont
  {Hochstrat}}, \bibinfo {author} {\bibfnamefont {C.}~\bibnamefont {Binek}},
  \bibinfo {author} {\bibfnamefont {X.}~\bibnamefont {Chen}}, \ and\ \bibinfo
  {author} {\bibfnamefont {W.}~\bibnamefont {Kleemann}},\ }\href {\doibase
  10.1016/j.jmmm.2003.12.344} {\bibfield  {journal} {\bibinfo  {journal} {J.
  Magn. Magn. Mater.}\ }\textbf {\bibinfo {volume} {272--276}},\ \bibinfo
  {pages} {325} (\bibinfo {year} {2004})}\BibitemShut {NoStop}%
\bibitem [{\citenamefont {Chu}\ \emph {et~al.}(2008)\citenamefont {Chu},
  \citenamefont {Martin}, \citenamefont {Holcomb}, \citenamefont {Gajek},
  \citenamefont {Han}, \citenamefont {He}, \citenamefont {Balke}, \citenamefont
  {Yang}, \citenamefont {Lee}, \citenamefont {Hu}, \citenamefont {Zhan},
  \citenamefont {Yang}, \citenamefont {Fraile-Rodr\'iguez}, \citenamefont
  {Scholl}, \citenamefont {Wang},\ and\ \citenamefont {Ramesh}}]{Chu:7:2008}%
  \BibitemOpen
  \bibfield  {author} {\bibinfo {author} {\bibfnamefont {Y.-H.}\ \bibnamefont
  {Chu}}, \bibinfo {author} {\bibfnamefont {L.~W.}\ \bibnamefont {Martin}},
  \bibinfo {author} {\bibfnamefont {M.~B.}\ \bibnamefont {Holcomb}}, \bibinfo
  {author} {\bibfnamefont {M.}~\bibnamefont {Gajek}}, \bibinfo {author}
  {\bibfnamefont {S.-J.}\ \bibnamefont {Han}}, \bibinfo {author} {\bibfnamefont
  {Q.}~\bibnamefont {He}}, \bibinfo {author} {\bibfnamefont {N.}~\bibnamefont
  {Balke}}, \bibinfo {author} {\bibfnamefont {C.-H.}\ \bibnamefont {Yang}},
  \bibinfo {author} {\bibfnamefont {D.}~\bibnamefont {Lee}}, \bibinfo {author}
  {\bibfnamefont {W.}~\bibnamefont {Hu}}, \bibinfo {author} {\bibfnamefont
  {Q.}~\bibnamefont {Zhan}}, \bibinfo {author} {\bibfnamefont {P.-L.}\
  \bibnamefont {Yang}}, \bibinfo {author} {\bibfnamefont {A.}~\bibnamefont
  {Fraile-Rodr\'iguez}}, \bibinfo {author} {\bibfnamefont {A.}~\bibnamefont
  {Scholl}}, \bibinfo {author} {\bibfnamefont {S.~X.}\ \bibnamefont {Wang}}, \
  and\ \bibinfo {author} {\bibfnamefont {R.}~\bibnamefont {Ramesh}},\ }\href
  {\doibase 10.1038/nmat2184} {\bibfield  {journal} {\bibinfo  {journal} {Nat.
  Mater.}\ }\textbf {\bibinfo {volume} {7}},\ \bibinfo {pages} {478} (\bibinfo
  {year} {2008})}\BibitemShut {NoStop}%
\bibitem [{\citenamefont {van Suchtelen}(1972)}]{Suchtelen:27:1972}%
  \BibitemOpen
  \bibfield  {author} {\bibinfo {author} {\bibfnamefont {J.}~\bibnamefont {van
  Suchtelen}},\ }\href@noop {} {\bibfield  {journal} {\bibinfo  {journal}
  {Philips Res. Rep.}\ }\textbf {\bibinfo {volume} {27}},\ \bibinfo {pages}
  {28} (\bibinfo {year} {1972})}\BibitemShut {NoStop}%
\bibitem [{\citenamefont {Thiele}\ \emph {et~al.}(2007)\citenamefont {Thiele},
  \citenamefont {D\"orr}, \citenamefont {Bilani}, \citenamefont {R\"odel},\
  and\ \citenamefont {Schultz}}]{Thiele:75:2007}%
  \BibitemOpen
  \bibfield  {author} {\bibinfo {author} {\bibfnamefont {C.}~\bibnamefont
  {Thiele}}, \bibinfo {author} {\bibfnamefont {K.}~\bibnamefont {D\"orr}},
  \bibinfo {author} {\bibfnamefont {O.}~\bibnamefont {Bilani}}, \bibinfo
  {author} {\bibfnamefont {J.}~\bibnamefont {R\"odel}}, \ and\ \bibinfo
  {author} {\bibfnamefont {L.}~\bibnamefont {Schultz}},\ }\href {\doibase
  10.1103/PhysRevB.75.054408} {\bibfield  {journal} {\bibinfo  {journal} {Phys.
  Rev. B}\ }\textbf {\bibinfo {volume} {75}},\ \bibinfo {pages} {054408}
  (\bibinfo {year} {2007})}\BibitemShut {NoStop}%
\bibitem [{\citenamefont {Eerenstein}\ \emph {et~al.}(2007)\citenamefont
  {Eerenstein}, \citenamefont {Wiora}, \citenamefont {Prieto}, \citenamefont
  {Scott},\ and\ \citenamefont {Mathur}}]{Eerenstein:6:2007}%
  \BibitemOpen
  \bibfield  {author} {\bibinfo {author} {\bibfnamefont {W.}~\bibnamefont
  {Eerenstein}}, \bibinfo {author} {\bibfnamefont {M.}~\bibnamefont {Wiora}},
  \bibinfo {author} {\bibfnamefont {J.~L.}\ \bibnamefont {Prieto}}, \bibinfo
  {author} {\bibfnamefont {J.~F.}\ \bibnamefont {Scott}}, \ and\ \bibinfo
  {author} {\bibfnamefont {N.~D.}\ \bibnamefont {Mathur}},\ }\href {\doibase
  10.1038/nmat1886} {\bibfield  {journal} {\bibinfo  {journal} {Nat. Mater.}\
  }\textbf {\bibinfo {volume} {6}},\ \bibinfo {pages} {348} (\bibinfo {year}
  {2007})}\BibitemShut {NoStop}%
\bibitem [{\citenamefont {Run}\ \emph {et~al.}(1974)\citenamefont {Run},
  \citenamefont {Terrell},\ and\ \citenamefont {Scholing}}]{Run:9:1974}%
  \BibitemOpen
  \bibfield  {author} {\bibinfo {author} {\bibfnamefont {A.~M. J.~G.}\
  \bibnamefont {Run}}, \bibinfo {author} {\bibfnamefont {D.~R.}\ \bibnamefont
  {Terrell}}, \ and\ \bibinfo {author} {\bibfnamefont {J.~H.}\ \bibnamefont
  {Scholing}},\ }\href {\doibase 10.1007/BF00540771} {\bibfield  {journal}
  {\bibinfo  {journal} {J. Mater. Sci.}\ }\textbf {\bibinfo {volume} {9}},\
  \bibinfo {pages} {1710} (\bibinfo {year} {1974})}\BibitemShut {NoStop}%
\bibitem [{\citenamefont {Ryu}\ \emph {et~al.}(2001)\citenamefont {Ryu},
  \citenamefont {Carazo}, \citenamefont {Uchino},\ and\ \citenamefont
  {Kim}}]{Ryu:40:2001}%
  \BibitemOpen
  \bibfield  {author} {\bibinfo {author} {\bibfnamefont {J.}~\bibnamefont
  {Ryu}}, \bibinfo {author} {\bibfnamefont {A.~V.}\ \bibnamefont {Carazo}},
  \bibinfo {author} {\bibfnamefont {K.}~\bibnamefont {Uchino}}, \ and\ \bibinfo
  {author} {\bibfnamefont {H.-E.}\ \bibnamefont {Kim}},\ }\href {\doibase
  10.1143/JJAP.40.4948} {\bibfield  {journal} {\bibinfo  {journal} {Jpn. J.
  Appl. Phys.}\ }\textbf {\bibinfo {volume} {40}},\ \bibinfo {pages} {4948}
  (\bibinfo {year} {2001})}\BibitemShut {NoStop}%
\bibitem [{\citenamefont {Zheng}\ \emph {et~al.}(2004)\citenamefont {Zheng},
  \citenamefont {Wang}, \citenamefont {Lofland}, \citenamefont {Ma},
  \citenamefont {{Mohaddes-Ardabili}}, \citenamefont {Zhao}, \citenamefont
  {{Salamanca-Riba}}, \citenamefont {Shinde}, \citenamefont {Ogale},
  \citenamefont {Bai}, \citenamefont {Viehland}, \citenamefont {Jia},
  \citenamefont {Schlom}, \citenamefont {Wuttig}, \citenamefont {Roytburd},\
  and\ \citenamefont {Ramesh}}]{Zheng:303:2004}%
  \BibitemOpen
  \bibfield  {author} {\bibinfo {author} {\bibfnamefont {H.}~\bibnamefont
  {Zheng}}, \bibinfo {author} {\bibfnamefont {J.}~\bibnamefont {Wang}},
  \bibinfo {author} {\bibfnamefont {S.~E.}\ \bibnamefont {Lofland}}, \bibinfo
  {author} {\bibfnamefont {Z.}~\bibnamefont {Ma}}, \bibinfo {author}
  {\bibfnamefont {L.}~\bibnamefont {{Mohaddes-Ardabili}}}, \bibinfo {author}
  {\bibfnamefont {T.}~\bibnamefont {Zhao}}, \bibinfo {author} {\bibfnamefont
  {L.}~\bibnamefont {{Salamanca-Riba}}}, \bibinfo {author} {\bibfnamefont
  {S.~R.}\ \bibnamefont {Shinde}}, \bibinfo {author} {\bibfnamefont {S.~B.}\
  \bibnamefont {Ogale}}, \bibinfo {author} {\bibfnamefont {F.}~\bibnamefont
  {Bai}}, \bibinfo {author} {\bibfnamefont {D.}~\bibnamefont {Viehland}},
  \bibinfo {author} {\bibfnamefont {Y.}~\bibnamefont {Jia}}, \bibinfo {author}
  {\bibfnamefont {D.~G.}\ \bibnamefont {Schlom}}, \bibinfo {author}
  {\bibfnamefont {M.}~\bibnamefont {Wuttig}}, \bibinfo {author} {\bibfnamefont
  {A.}~\bibnamefont {Roytburd}}, \ and\ \bibinfo {author} {\bibfnamefont
  {R.}~\bibnamefont {Ramesh}},\ }\href {\doibase 10.1126/science.1094207}
  {\bibfield  {journal} {\bibinfo  {journal} {Science}\ }\textbf {\bibinfo
  {volume} {303}},\ \bibinfo {pages} {661} (\bibinfo {year}
  {2004})}\BibitemShut {NoStop}%
\bibitem [{\citenamefont {Vaz}\ \emph {et~al.}(2009)\citenamefont {Vaz},
  \citenamefont {Hoffman}, \citenamefont {Posadas},\ and\ \citenamefont
  {Ahn}}]{Vaz:94:2009}%
  \BibitemOpen
  \bibfield  {author} {\bibinfo {author} {\bibfnamefont {C.~A.~F.}\
  \bibnamefont {Vaz}}, \bibinfo {author} {\bibfnamefont {J.}~\bibnamefont
  {Hoffman}}, \bibinfo {author} {\bibfnamefont {A.-B.}\ \bibnamefont
  {Posadas}}, \ and\ \bibinfo {author} {\bibfnamefont {C.~H.}\ \bibnamefont
  {Ahn}},\ }\href {\doibase 10.1063/1.3069280} {\bibfield  {journal} {\bibinfo
  {journal} {Appl. Phys. Lett.}\ }\textbf {\bibinfo {volume} {94}},\ \bibinfo
  {pages} {022504} (\bibinfo {year} {2009})}\BibitemShut {NoStop}%
\bibitem [{\citenamefont {Ziese}\ \emph {et~al.}(2006)\citenamefont {Ziese},
  \citenamefont {Bollero}, \citenamefont {Panagiotopoulos},\ and\ \citenamefont
  {Moutis}}]{Ziese:88:2006}%
  \BibitemOpen
  \bibfield  {author} {\bibinfo {author} {\bibfnamefont {M.}~\bibnamefont
  {Ziese}}, \bibinfo {author} {\bibfnamefont {A.}~\bibnamefont {Bollero}},
  \bibinfo {author} {\bibfnamefont {I.}~\bibnamefont {Panagiotopoulos}}, \ and\
  \bibinfo {author} {\bibfnamefont {N.}~\bibnamefont {Moutis}},\ }\href
  {\doibase 10.1063/1.2206121} {\bibfield  {journal} {\bibinfo  {journal}
  {Appl. Phys. Lett.}\ }\textbf {\bibinfo {volume} {88}},\ \bibinfo {pages}
  {212502} (\bibinfo {year} {2006})}\BibitemShut {NoStop}%
\bibitem [{\citenamefont {Tian}\ \emph {et~al.}(2008)\citenamefont {Tian},
  \citenamefont {Qu}, \citenamefont {Luo}, \citenamefont {Yang}, \citenamefont
  {Guo}, \citenamefont {Zhang}, \citenamefont {Zhao},\ and\ \citenamefont
  {Li}}]{Tian:92:2008}%
  \BibitemOpen
  \bibfield  {author} {\bibinfo {author} {\bibfnamefont {H.~F.}\ \bibnamefont
  {Tian}}, \bibinfo {author} {\bibfnamefont {T.~L.}\ \bibnamefont {Qu}},
  \bibinfo {author} {\bibfnamefont {L.~B.}\ \bibnamefont {Luo}}, \bibinfo
  {author} {\bibfnamefont {J.~J.}\ \bibnamefont {Yang}}, \bibinfo {author}
  {\bibfnamefont {S.~M.}\ \bibnamefont {Guo}}, \bibinfo {author} {\bibfnamefont
  {H.~Y.}\ \bibnamefont {Zhang}}, \bibinfo {author} {\bibfnamefont {Y.~G.}\
  \bibnamefont {Zhao}}, \ and\ \bibinfo {author} {\bibfnamefont {J.~Q.}\
  \bibnamefont {Li}},\ }\href {\doibase 10.1063/1.2844858} {\bibfield
  {journal} {\bibinfo  {journal} {Appl. Phys. Lett.}\ }\textbf {\bibinfo
  {volume} {92}},\ \bibinfo {pages} {063507} (\bibinfo {year}
  {2008})}\BibitemShut {NoStop}%
\bibitem [{\citenamefont {Sterbinsky}\ \emph {et~al.}(2010)\citenamefont
  {Sterbinsky}, \citenamefont {Wessels}, \citenamefont {Kim}, \citenamefont
  {Karapetrova}, \citenamefont {Ryan},\ and\ \citenamefont
  {Keavney}}]{Sterbinsky:96:2010}%
  \BibitemOpen
  \bibfield  {author} {\bibinfo {author} {\bibfnamefont {G.~E.}\ \bibnamefont
  {Sterbinsky}}, \bibinfo {author} {\bibfnamefont {B.~W.}\ \bibnamefont
  {Wessels}}, \bibinfo {author} {\bibfnamefont {J.-W.}\ \bibnamefont {Kim}},
  \bibinfo {author} {\bibfnamefont {E.}~\bibnamefont {Karapetrova}}, \bibinfo
  {author} {\bibfnamefont {P.~J.}\ \bibnamefont {Ryan}}, \ and\ \bibinfo
  {author} {\bibfnamefont {D.~J.}\ \bibnamefont {Keavney}},\ }\href {\doibase
  10.1063/1.3330890} {\bibfield  {journal} {\bibinfo  {journal} {Appl. Phys.
  Lett.}\ }\textbf {\bibinfo {volume} {96}},\ \bibinfo {pages} {092510}
  (\bibinfo {year} {2010})}\BibitemShut {NoStop}%
\bibitem [{\citenamefont {Koo}\ \emph {et~al.}(2009)\citenamefont {Koo},
  \citenamefont {Song}, \citenamefont {Hur}, \citenamefont {Jung},
  \citenamefont {Jang}, \citenamefont {Lee}, \citenamefont {Koo}, \citenamefont
  {Jeong}, \citenamefont {Cho},\ and\ \citenamefont {Jo}}]{Koo:94:2009}%
  \BibitemOpen
  \bibfield  {author} {\bibinfo {author} {\bibfnamefont {Y.~S.}\ \bibnamefont
  {Koo}}, \bibinfo {author} {\bibfnamefont {K.~M.}\ \bibnamefont {Song}},
  \bibinfo {author} {\bibfnamefont {N.}~\bibnamefont {Hur}}, \bibinfo {author}
  {\bibfnamefont {J.~H.}\ \bibnamefont {Jung}}, \bibinfo {author}
  {\bibfnamefont {T.-H.}\ \bibnamefont {Jang}}, \bibinfo {author}
  {\bibfnamefont {H.~J.}\ \bibnamefont {Lee}}, \bibinfo {author} {\bibfnamefont
  {T.~Y.}\ \bibnamefont {Koo}}, \bibinfo {author} {\bibfnamefont {Y.~H.}\
  \bibnamefont {Jeong}}, \bibinfo {author} {\bibfnamefont {J.~H.}\ \bibnamefont
  {Cho}}, \ and\ \bibinfo {author} {\bibfnamefont {Y.~H.}\ \bibnamefont {Jo}},\
  }\href {\doibase 10.1063/1.3073751} {\bibfield  {journal} {\bibinfo
  {journal} {Appl. Phys. Lett.}\ }\textbf {\bibinfo {volume} {94}},\ \bibinfo
  {pages} {032903} (\bibinfo {year} {2009})}\BibitemShut {NoStop}%
\bibitem [{\citenamefont {Niranjan}\ \emph {et~al.}(2008)\citenamefont
  {Niranjan}, \citenamefont {Velev}, \citenamefont {Duan}, \citenamefont
  {Jaswal},\ and\ \citenamefont {Tsymbal}}]{Niranjan:PRB:78:2008}%
  \BibitemOpen
  \bibfield  {author} {\bibinfo {author} {\bibfnamefont {M.~K.}\ \bibnamefont
  {Niranjan}}, \bibinfo {author} {\bibfnamefont {J.~P.}\ \bibnamefont {Velev}},
  \bibinfo {author} {\bibfnamefont {C.-G.}\ \bibnamefont {Duan}}, \bibinfo
  {author} {\bibfnamefont {S.~S.}\ \bibnamefont {Jaswal}}, \ and\ \bibinfo
  {author} {\bibfnamefont {E.~Y.}\ \bibnamefont {Tsymbal}},\ }\href {\doibase
  10.1103/PhysRevB.78.104405} {\bibfield  {journal} {\bibinfo  {journal} {Phys.
  Rev. B}\ }\textbf {\bibinfo {volume} {78}},\ \bibinfo {pages} {104405}
  (\bibinfo {year} {2008})}\BibitemShut {NoStop}%
\bibitem [{\citenamefont {Chopdekar}\ and\ \citenamefont
  {Suzuki}(2006)}]{Chopdekar:89:2006}%
  \BibitemOpen
  \bibfield  {author} {\bibinfo {author} {\bibfnamefont {R.~V.}\ \bibnamefont
  {Chopdekar}}\ and\ \bibinfo {author} {\bibfnamefont {Y.}~\bibnamefont
  {Suzuki}},\ }\href {\doibase 10.1063/1.2370881} {\bibfield  {journal}
  {\bibinfo  {journal} {Appl. Phys. Lett.}\ }\textbf {\bibinfo {volume} {89}},\
  \bibinfo {pages} {182506} (\bibinfo {year} {2006})}\BibitemShut {NoStop}%
\bibitem [{\citenamefont {Pan}\ \emph {et~al.}(2013)\citenamefont {Pan},
  \citenamefont {Hong}, \citenamefont {Guest}, \citenamefont {Liu},\ and\
  \citenamefont {Petford-Long}}]{Pan:46:2013}%
  \BibitemOpen
  \bibfield  {author} {\bibinfo {author} {\bibfnamefont {M.}~\bibnamefont
  {Pan}}, \bibinfo {author} {\bibfnamefont {S.}~\bibnamefont {Hong}}, \bibinfo
  {author} {\bibfnamefont {J.~R.}\ \bibnamefont {Guest}}, \bibinfo {author}
  {\bibfnamefont {Y.}~\bibnamefont {Liu}}, \ and\ \bibinfo {author}
  {\bibfnamefont {A.}~\bibnamefont {Petford-Long}},\ }\href {\doibase
  10.1088/0022-3727/46/5/055001} {\bibfield  {journal} {\bibinfo  {journal} {J.
  Phys. D: Appl. Phys.}\ }\textbf {\bibinfo {volume} {46}},\ \bibinfo {pages}
  {055001} (\bibinfo {year} {2013})}\BibitemShut {NoStop}%
\bibitem [{\citenamefont {Chopdekar}\ \emph {et~al.}(2012)\citenamefont
  {Chopdekar}, \citenamefont {Malik}, \citenamefont {Fraile~Rodr\'iguez},
  \citenamefont {Le~Guyader}, \citenamefont {Takamura}, \citenamefont {Scholl},
  \citenamefont {Stender}, \citenamefont {Schneider}, \citenamefont {Bernhard},
  \citenamefont {Nolting},\ and\ \citenamefont
  {Heyderman}}]{Chopdekar:86:2012}%
  \BibitemOpen
  \bibfield  {author} {\bibinfo {author} {\bibfnamefont {R.~V.}\ \bibnamefont
  {Chopdekar}}, \bibinfo {author} {\bibfnamefont {V.~K.}\ \bibnamefont
  {Malik}}, \bibinfo {author} {\bibfnamefont {A.}~\bibnamefont
  {Fraile~Rodr\'iguez}}, \bibinfo {author} {\bibfnamefont {L.}~\bibnamefont
  {Le~Guyader}}, \bibinfo {author} {\bibfnamefont {Y.}~\bibnamefont
  {Takamura}}, \bibinfo {author} {\bibfnamefont {A.}~\bibnamefont {Scholl}},
  \bibinfo {author} {\bibfnamefont {D.}~\bibnamefont {Stender}}, \bibinfo
  {author} {\bibfnamefont {C.~W.}\ \bibnamefont {Schneider}}, \bibinfo {author}
  {\bibfnamefont {C.}~\bibnamefont {Bernhard}}, \bibinfo {author}
  {\bibfnamefont {F.}~\bibnamefont {Nolting}}, \ and\ \bibinfo {author}
  {\bibfnamefont {L.~J.}\ \bibnamefont {Heyderman}},\ }\href {\doibase
  10.1103/PhysRevB.86.014408} {\bibfield  {journal} {\bibinfo  {journal} {Phys.
  Rev. B}\ }\textbf {\bibinfo {volume} {86}},\ \bibinfo {pages} {014408}
  (\bibinfo {year} {2012})}\BibitemShut {NoStop}%
\bibitem [{\citenamefont {Dale}\ \emph {et~al.}(2003)\citenamefont {Dale},
  \citenamefont {Fleet}, \citenamefont {Brock},\ and\ \citenamefont
  {Suzuki}}]{Dale:82:2003}%
  \BibitemOpen
  \bibfield  {author} {\bibinfo {author} {\bibfnamefont {D.}~\bibnamefont
  {Dale}}, \bibinfo {author} {\bibfnamefont {A.}~\bibnamefont {Fleet}},
  \bibinfo {author} {\bibfnamefont {J.~D.}\ \bibnamefont {Brock}}, \ and\
  \bibinfo {author} {\bibfnamefont {Y.}~\bibnamefont {Suzuki}},\ }\href
  {\doibase 10.1063/1.1578186} {\bibfield  {journal} {\bibinfo  {journal}
  {Appl. Phys. Lett.}\ }\textbf {\bibinfo {volume} {82}},\ \bibinfo {pages}
  {3725} (\bibinfo {year} {2003})}\BibitemShut {NoStop}%
\bibitem [{\citenamefont {Singh}\ \emph {et~al.}(2006)\citenamefont {Singh},
  \citenamefont {Prellier}, \citenamefont {Mechin}, \citenamefont {Simon},\
  and\ \citenamefont {Raveau}}]{Singh:99:2006}%
  \BibitemOpen
  \bibfield  {author} {\bibinfo {author} {\bibfnamefont {M.~P.}\ \bibnamefont
  {Singh}}, \bibinfo {author} {\bibfnamefont {W.}~\bibnamefont {Prellier}},
  \bibinfo {author} {\bibfnamefont {L.}~\bibnamefont {Mechin}}, \bibinfo
  {author} {\bibfnamefont {C.}~\bibnamefont {Simon}}, \ and\ \bibinfo {author}
  {\bibfnamefont {B.}~\bibnamefont {Raveau}},\ }\href {\doibase
  10.1063/1.2161424} {\bibfield  {journal} {\bibinfo  {journal} {J. Appl.
  Phys.}\ }\textbf {\bibinfo {volume} {99}},\ \bibinfo {pages} {024105}
  (\bibinfo {year} {2006})}\BibitemShut {NoStop}%
\bibitem [{\citenamefont {Alberca}\ \emph {et~al.}(2012)\citenamefont
  {Alberca}, \citenamefont {Munuera}, \citenamefont {Tornos}, \citenamefont
  {Mompean}, \citenamefont {Biskup}, \citenamefont {Ruiz}, \citenamefont
  {Nemes}, \citenamefont {de~Andres}, \citenamefont {Le\'on}, \citenamefont
  {Santamar\'ia},\ and\ \citenamefont
  {Garc\'ia-Hern\'andez}}]{Alberca:86:2012}%
  \BibitemOpen
  \bibfield  {author} {\bibinfo {author} {\bibfnamefont {A.}~\bibnamefont
  {Alberca}}, \bibinfo {author} {\bibfnamefont {C.}~\bibnamefont {Munuera}},
  \bibinfo {author} {\bibfnamefont {J.}~\bibnamefont {Tornos}}, \bibinfo
  {author} {\bibfnamefont {F.~J.}\ \bibnamefont {Mompean}}, \bibinfo {author}
  {\bibfnamefont {N.}~\bibnamefont {Biskup}}, \bibinfo {author} {\bibfnamefont
  {A.}~\bibnamefont {Ruiz}}, \bibinfo {author} {\bibfnamefont {N.~M.}\
  \bibnamefont {Nemes}}, \bibinfo {author} {\bibfnamefont {A.}~\bibnamefont
  {de~Andres}}, \bibinfo {author} {\bibfnamefont {C.}~\bibnamefont {Le\'on}},
  \bibinfo {author} {\bibfnamefont {J.}~\bibnamefont
  {Santamar\'ia}}, \ and\ \bibinfo {author} {\bibfnamefont {M.}~\bibnamefont
  {Garc\'ia-Hern\'andez}},\ }\href {\doibase 10.1103/PhysRevB.86.144416}
  {\bibfield  {journal} {\bibinfo  {journal} {Phys. Rev. B}\ }\textbf {\bibinfo
  {volume} {86}},\ \bibinfo {pages} {144416} (\bibinfo {year}
  {2012})}\BibitemShut {NoStop}%
\bibitem [{\citenamefont {Murugavel}\ \emph {et~al.}(2004)\citenamefont
  {Murugavel}, \citenamefont {Padhan},\ and\ \citenamefont
  {Prellier}}]{Murugavel:85:2004}%
  \BibitemOpen
  \bibfield  {author} {\bibinfo {author} {\bibfnamefont {P.}~\bibnamefont
  {Murugavel}}, \bibinfo {author} {\bibfnamefont {P.}~\bibnamefont {Padhan}}, \
  and\ \bibinfo {author} {\bibfnamefont {W.}~\bibnamefont {Prellier}},\ }\href
  {\doibase 10.1063/1.1825075} {\bibfield  {journal} {\bibinfo  {journal}
  {Appl. Phys. Lett.}\ }\textbf {\bibinfo {volume} {85}},\ \bibinfo {pages}
  {4992} (\bibinfo {year} {2004})}\BibitemShut {NoStop}%
\bibitem [{\citenamefont {Czeschka}\ \emph {et~al.}(2009)\citenamefont
  {Czeschka}, \citenamefont {Gepr\"{a}gs}, \citenamefont {Opel}, \citenamefont
  {Goennenwein},\ and\ \citenamefont {Gross}}]{Czeschka:95:2009}%
  \BibitemOpen
  \bibfield  {author} {\bibinfo {author} {\bibfnamefont {F.~D.}\ \bibnamefont
  {Czeschka}}, \bibinfo {author} {\bibfnamefont {S.}~\bibnamefont
  {Gepr\"{a}gs}}, \bibinfo {author} {\bibfnamefont {M.}~\bibnamefont {Opel}},
  \bibinfo {author} {\bibfnamefont {S.~T.~B.}\ \bibnamefont {Goennenwein}}, \
  and\ \bibinfo {author} {\bibfnamefont {R.}~\bibnamefont {Gross}},\ }\href
  {\doibase 10.1063/1.3200236} {\bibfield  {journal} {\bibinfo  {journal}
  {Appl. Phys. Lett.}\ }\textbf {\bibinfo {volume} {95}},\ \bibinfo {pages}
  {062508} (\bibinfo {year} {2009})}\BibitemShut {NoStop}%
\bibitem [{\citenamefont {Komelj}(2010)}]{Komelj:82:2010}%
  \BibitemOpen
  \bibfield  {author} {\bibinfo {author} {\bibfnamefont {M.}~\bibnamefont
  {Komelj}},\ }\href {\doibase 10.1103/PhysRevB.82.012410} {\bibfield
  {journal} {\bibinfo  {journal} {Phys. Rev. B}\ }\textbf {\bibinfo {volume}
  {82}},\ \bibinfo {pages} {012410} (\bibinfo {year} {2010})}\BibitemShut
  {NoStop}%
\bibitem [{\citenamefont {Opel}\ \emph {et~al.}(2011)\citenamefont {Opel},
  \citenamefont {Gepr\"{a}gs}, \citenamefont {Menzel}, \citenamefont {Nielsen},
  \citenamefont {Reisinger}, \citenamefont {Nielsen}, \citenamefont
  {Brandlmaier}, \citenamefont {Czeschka}, \citenamefont {Althammer},
  \citenamefont {Weiler}, \citenamefont {Goennenwein}, \citenamefont {Simon},
  \citenamefont {Svete}, \citenamefont {Yu}, \citenamefont {H\"{u}hne},
  \citenamefont {Mader},\ and\ \citenamefont {Gross}}]{Opel:208:2011}%
  \BibitemOpen
  \bibfield  {author} {\bibinfo {author} {\bibfnamefont {M.}~\bibnamefont
  {Opel}}, \bibinfo {author} {\bibfnamefont {S.}~\bibnamefont {Gepr\"{a}gs}},
  \bibinfo {author} {\bibfnamefont {E.~P.}\ \bibnamefont {Menzel}}, \bibinfo
  {author} {\bibfnamefont {A.}~\bibnamefont {Nielsen}}, \bibinfo {author}
  {\bibfnamefont {D.}~\bibnamefont {Reisinger}}, \bibinfo {author}
  {\bibfnamefont {K.}~\bibnamefont {Nielsen}}, \bibinfo {author} {\bibfnamefont
  {A.}~\bibnamefont {Brandlmaier}}, \bibinfo {author} {\bibfnamefont {F.~D.}\
  \bibnamefont {Czeschka}}, \bibinfo {author} {\bibfnamefont {M.}~\bibnamefont
  {Althammer}}, \bibinfo {author} {\bibfnamefont {M.}~\bibnamefont {Weiler}},
  \bibinfo {author} {\bibfnamefont {S.~T.~B.}\ \bibnamefont {Goennenwein}},
  \bibinfo {author} {\bibfnamefont {J.}~\bibnamefont {Simon}}, \bibinfo
  {author} {\bibfnamefont {M.}~\bibnamefont {Svete}}, \bibinfo {author}
  {\bibfnamefont {W.}~\bibnamefont {Yu}}, \bibinfo {author} {\bibfnamefont
  {S.}~\bibnamefont {H\"{u}hne}}, \bibinfo {author} {\bibfnamefont
  {W.}~\bibnamefont {Mader}}, \ and\ \bibinfo {author} {\bibfnamefont
  {R.}~\bibnamefont {Gross}},\ }\href {\doibase 10.1002/pssa.201026403}
  {\bibfield  {journal} {\bibinfo  {journal} {Phys. Status Solidi A}\ }\textbf
  {\bibinfo {volume} {208}},\ \bibinfo {pages} {232} (\bibinfo {year}
  {2011})}\BibitemShut {NoStop}%
\bibitem [{\citenamefont {Shirahata}\ \emph {et~al.}(2011)\citenamefont
  {Shirahata}, \citenamefont {Nozaki}, \citenamefont {Venkataiah},
  \citenamefont {Taniguchi}, \citenamefont {Itoh},\ and\ \citenamefont
  {Taniyama}}]{Shirahata:99:2011}%
  \BibitemOpen
  \bibfield  {author} {\bibinfo {author} {\bibfnamefont {Y.}~\bibnamefont
  {Shirahata}}, \bibinfo {author} {\bibfnamefont {T.}~\bibnamefont {Nozaki}},
  \bibinfo {author} {\bibfnamefont {G.}~\bibnamefont {Venkataiah}}, \bibinfo
  {author} {\bibfnamefont {H.}~\bibnamefont {Taniguchi}}, \bibinfo {author}
  {\bibfnamefont {M.}~\bibnamefont {Itoh}}, \ and\ \bibinfo {author}
  {\bibfnamefont {T.}~\bibnamefont {Taniyama}},\ }\href {\doibase
  10.1063/1.3609237} {\bibfield  {journal} {\bibinfo  {journal} {Appl. Phys.
  Lett.}\ }\textbf {\bibinfo {volume} {99}},\ \bibinfo {pages} {022501}
  (\bibinfo {year} {2011})}\BibitemShut {NoStop}%
\bibitem [{\citenamefont {Venkataiah}\ \emph {et~al.}(2012)\citenamefont
  {Venkataiah}, \citenamefont {Shirahata}, \citenamefont {Suzuki},
  \citenamefont {Itoh},\ and\ \citenamefont {Taniyama}}]{Venkataiah:111:2012}%
  \BibitemOpen
  \bibfield  {author} {\bibinfo {author} {\bibfnamefont {G.}~\bibnamefont
  {Venkataiah}}, \bibinfo {author} {\bibfnamefont {Y.}~\bibnamefont
  {Shirahata}}, \bibinfo {author} {\bibfnamefont {I.}~\bibnamefont {Suzuki}},
  \bibinfo {author} {\bibfnamefont {M.}~\bibnamefont {Itoh}}, \ and\ \bibinfo
  {author} {\bibfnamefont {T.}~\bibnamefont {Taniyama}},\ }\href {\doibase
  10.1063/1.3684695} {\bibfield  {journal} {\bibinfo  {journal} {J. Appl.
  Phys.}\ }\textbf {\bibinfo {volume} {111}},\ \bibinfo {pages} {033921}
  (\bibinfo {year} {2012})}\BibitemShut {NoStop}%
\bibitem [{\citenamefont {Sahoo}\ \emph {et~al.}(2007)\citenamefont {Sahoo},
  \citenamefont {Polisetty}, \citenamefont {Duan}, \citenamefont {Jaswal},
  \citenamefont {Tsymbal},\ and\ \citenamefont {Binek}}]{Sahoo:76:2007}%
  \BibitemOpen
  \bibfield  {author} {\bibinfo {author} {\bibfnamefont {S.}~\bibnamefont
  {Sahoo}}, \bibinfo {author} {\bibfnamefont {S.}~\bibnamefont {Polisetty}},
  \bibinfo {author} {\bibfnamefont {C.-G.}\ \bibnamefont {Duan}}, \bibinfo
  {author} {\bibfnamefont {S.~S.}\ \bibnamefont {Jaswal}}, \bibinfo {author}
  {\bibfnamefont {E.~Y.}\ \bibnamefont {Tsymbal}}, \ and\ \bibinfo {author}
  {\bibfnamefont {C.}~\bibnamefont {Binek}},\ }\href {\doibase
  10.1103/PhysRevB.76.092108} {\bibfield  {journal} {\bibinfo  {journal} {Phys.
  Rev. B}\ }\textbf {\bibinfo {volume} {76}},\ \bibinfo {pages} {092108}
  (\bibinfo {year} {2007})}\BibitemShut {NoStop}%
\bibitem [{\citenamefont {Venkataiah}\ \emph {et~al.}(2011)\citenamefont
  {Venkataiah}, \citenamefont {Shirahata}, \citenamefont {Itoh},\ and\
  \citenamefont {Taniyama}}]{Venkataiah:99:2011}%
  \BibitemOpen
  \bibfield  {author} {\bibinfo {author} {\bibfnamefont {G.}~\bibnamefont
  {Venkataiah}}, \bibinfo {author} {\bibfnamefont {Y.}~\bibnamefont
  {Shirahata}}, \bibinfo {author} {\bibfnamefont {M.}~\bibnamefont {Itoh}}, \
  and\ \bibinfo {author} {\bibfnamefont {T.}~\bibnamefont {Taniyama}},\ }\href
  {\doibase 10.1063/1.3628464} {\bibfield  {journal} {\bibinfo  {journal}
  {Appl. Phys. Lett.}\ }\textbf {\bibinfo {volume} {99}},\ \bibinfo {pages}
  {102506} (\bibinfo {year} {2011})}\BibitemShut {NoStop}%
\bibitem [{\citenamefont {Gepr\"{a}gs}\ \emph {et~al.}(2010)\citenamefont
  {Gepr\"{a}gs}, \citenamefont {Brandlmaier}, \citenamefont {Opel},
  \citenamefont {Gross},\ and\ \citenamefont {Goennenwein}}]{Gepraegs:96:2010}%
  \BibitemOpen
  \bibfield  {author} {\bibinfo {author} {\bibfnamefont {S.}~\bibnamefont
  {Gepr\"{a}gs}}, \bibinfo {author} {\bibfnamefont {A.}~\bibnamefont
  {Brandlmaier}}, \bibinfo {author} {\bibfnamefont {M.}~\bibnamefont {Opel}},
  \bibinfo {author} {\bibfnamefont {R.}~\bibnamefont {Gross}}, \ and\ \bibinfo
  {author} {\bibfnamefont {S.~T.~B.}\ \bibnamefont {Goennenwein}},\ }\href
  {\doibase 10.1063/1.3377923} {\bibfield  {journal} {\bibinfo  {journal}
  {Appl. Phys. Lett.}\ }\textbf {\bibinfo {volume} {96}},\ \bibinfo {pages}
  {142509} (\bibinfo {year} {2010})}\BibitemShut {NoStop}%
\bibitem [{\citenamefont {Shu}\ \emph {et~al.}(2012)\citenamefont {Shu},
  \citenamefont {Li}, \citenamefont {Ma}, \citenamefont {Gao}, \citenamefont
  {Gu}, \citenamefont {Shen}, \citenamefont {Lin},\ and\ \citenamefont
  {Nan}}]{Shu:100:2012}%
  \BibitemOpen
  \bibfield  {author} {\bibinfo {author} {\bibfnamefont {L.}~\bibnamefont
  {Shu}}, \bibinfo {author} {\bibfnamefont {Z.}~\bibnamefont {Li}}, \bibinfo
  {author} {\bibfnamefont {J.}~\bibnamefont {Ma}}, \bibinfo {author}
  {\bibfnamefont {Y.}~\bibnamefont {Gao}}, \bibinfo {author} {\bibfnamefont
  {L.}~\bibnamefont {Gu}}, \bibinfo {author} {\bibfnamefont {Y.}~\bibnamefont
  {Shen}}, \bibinfo {author} {\bibfnamefont {Y.}~\bibnamefont {Lin}}, \ and\
  \bibinfo {author} {\bibfnamefont {C.~W.}\ \bibnamefont {Nan}},\ }\href
  {\doibase 10.1063/1.3675868} {\bibfield  {journal} {\bibinfo  {journal}
  {Appl. Phys. Lett.}\ }\textbf {\bibinfo {volume} {100}},\ \bibinfo {pages}
  {022405} (\bibinfo {year} {2012})}\BibitemShut {NoStop}%
\bibitem [{\citenamefont {Ghidini}\ \emph {et~al.}(2013)\citenamefont
  {Ghidini}, \citenamefont {Pellicelli}, \citenamefont {Prieto}, \citenamefont
  {Moya}, \citenamefont {Soussi}, \citenamefont {Briscoe}, \citenamefont
  {Dunn},\ and\ \citenamefont {Mathur}}]{Ghidini:4:2013}%
  \BibitemOpen
  \bibfield  {author} {\bibinfo {author} {\bibfnamefont {M.}~\bibnamefont
  {Ghidini}}, \bibinfo {author} {\bibfnamefont {R.}~\bibnamefont {Pellicelli}},
  \bibinfo {author} {\bibfnamefont {J.~L.}\ \bibnamefont {Prieto}}, \bibinfo
  {author} {\bibfnamefont {X.}~\bibnamefont {Moya}}, \bibinfo {author}
  {\bibfnamefont {J.}~\bibnamefont {Soussi}}, \bibinfo {author} {\bibfnamefont
  {J.}~\bibnamefont {Briscoe}}, \bibinfo {author} {\bibfnamefont
  {S.}~\bibnamefont {Dunn}}, \ and\ \bibinfo {author} {\bibfnamefont {N.~D.}\
  \bibnamefont {Mathur}},\ }\href {\doibase 10.1038/ncomms2398} {\bibfield
  {journal} {\bibinfo  {journal} {Nature Comm.}\ }\textbf {\bibinfo {volume}
  {4}},\ \bibinfo {pages} {1453} (\bibinfo {year} {2013})}\BibitemShut
  {NoStop}%
\bibitem [{\citenamefont {Streubel}\ \emph {et~al.}(2013)\citenamefont
  {Streubel}, \citenamefont {K\"ohler}, \citenamefont {Sch\"afer},\ and\
  \citenamefont {Eng}}]{Streubel:87:2013}%
  \BibitemOpen
  \bibfield  {author} {\bibinfo {author} {\bibfnamefont {R.}~\bibnamefont
  {Streubel}}, \bibinfo {author} {\bibfnamefont {D.}~\bibnamefont {K\"ohler}},
  \bibinfo {author} {\bibfnamefont {R.}~\bibnamefont {Sch\"afer}}, \ and\
  \bibinfo {author} {\bibfnamefont {L.~M.}\ \bibnamefont {Eng}},\ }\href
  {\doibase 10.1103/PhysRevB.87.054410} {\bibfield  {journal} {\bibinfo
  {journal} {Phys. Rev. B}\ }\textbf {\bibinfo {volume} {87}},\ \bibinfo
  {pages} {054410} (\bibinfo {year} {2013})}\BibitemShut {NoStop}%
\bibitem [{\citenamefont {Narayanan}\ \emph {et~al.}(2012)\citenamefont
  {Narayanan}, \citenamefont {Mandal}, \citenamefont {Tyagi}, \citenamefont
  {Kumarasiri}, \citenamefont {Zhan}, \citenamefont {Hahm}, \citenamefont
  {Anantharaman}, \citenamefont {Lawes},\ and\ \citenamefont
  {Ajayan}}]{Narayanan:12:2012}%
  \BibitemOpen
  \bibfield  {author} {\bibinfo {author} {\bibfnamefont {T.~N.}\ \bibnamefont
  {Narayanan}}, \bibinfo {author} {\bibfnamefont {B.~P.}\ \bibnamefont
  {Mandal}}, \bibinfo {author} {\bibfnamefont {A.~K.}\ \bibnamefont {Tyagi}},
  \bibinfo {author} {\bibfnamefont {A.}~\bibnamefont {Kumarasiri}}, \bibinfo
  {author} {\bibfnamefont {X.}~\bibnamefont {Zhan}}, \bibinfo {author}
  {\bibfnamefont {M.~G.}\ \bibnamefont {Hahm}}, \bibinfo {author}
  {\bibfnamefont {M.~R.}\ \bibnamefont {Anantharaman}}, \bibinfo {author}
  {\bibfnamefont {G.}~\bibnamefont {Lawes}}, \ and\ \bibinfo {author}
  {\bibfnamefont {P.~M.}\ \bibnamefont {Ajayan}},\ }\href {\doibase
  10.1021/nl300849u} {\bibfield  {journal} {\bibinfo  {journal} {Nano Lett.}\
  }\textbf {\bibinfo {volume} {12}},\ \bibinfo {pages} {3025} (\bibinfo {year}
  {2012})}\BibitemShut {NoStop}%
\bibitem [{\citenamefont {Lahtinen}\ \emph {et~al.}(2012)\citenamefont
  {Lahtinen}, \citenamefont {Franke},\ and\ \citenamefont {van
  Dijken}}]{Lahtinen:2:2012}%
  \BibitemOpen
  \bibfield  {author} {\bibinfo {author} {\bibfnamefont {T.~H.~E.}\
  \bibnamefont {Lahtinen}}, \bibinfo {author} {\bibfnamefont {K.~J.~A.}\
  \bibnamefont {Franke}}, \ and\ \bibinfo {author} {\bibfnamefont
  {S.}~\bibnamefont {van Dijken}},\ }\href {\doibase 10.1038/srep00258}
  {\bibfield  {journal} {\bibinfo  {journal} {Sci. Rep.}\ }\textbf {\bibinfo
  {volume} {2}},\ \bibinfo {pages} {258} (\bibinfo {year} {2012})}\BibitemShut
  {NoStop}%
\bibitem [{\citenamefont {Brandlmaier}\ \emph {et~al.}(2008)\citenamefont
  {Brandlmaier}, \citenamefont {Gepr\"ags}, \citenamefont {Weiler},
  \citenamefont {Boger}, \citenamefont {Opel}, \citenamefont {Huebl},
  \citenamefont {Bihler}, \citenamefont {Brandt}, \citenamefont {Botters},
  \citenamefont {Grundler}, \citenamefont {Gross},\ and\ \citenamefont
  {Goennenwein}}]{Brandlmaier:77:2008}%
  \BibitemOpen
  \bibfield  {author} {\bibinfo {author} {\bibfnamefont {A.}~\bibnamefont
  {Brandlmaier}}, \bibinfo {author} {\bibfnamefont {S.}~\bibnamefont
  {Gepr\"ags}}, \bibinfo {author} {\bibfnamefont {M.}~\bibnamefont {Weiler}},
  \bibinfo {author} {\bibfnamefont {A.}~\bibnamefont {Boger}}, \bibinfo
  {author} {\bibfnamefont {M.}~\bibnamefont {Opel}}, \bibinfo {author}
  {\bibfnamefont {H.}~\bibnamefont {Huebl}}, \bibinfo {author} {\bibfnamefont
  {C.}~\bibnamefont {Bihler}}, \bibinfo {author} {\bibfnamefont {M.~S.}\
  \bibnamefont {Brandt}}, \bibinfo {author} {\bibfnamefont {B.}~\bibnamefont
  {Botters}}, \bibinfo {author} {\bibfnamefont {D.}~\bibnamefont {Grundler}},
  \bibinfo {author} {\bibfnamefont {R.}~\bibnamefont {Gross}}, \ and\ \bibinfo
  {author} {\bibfnamefont {S.~T.~B.}\ \bibnamefont {Goennenwein}},\ }\href
  {\doibase 10.1103/PhysRevB.77.104445} {\bibfield  {journal} {\bibinfo
  {journal} {Phys. Rev. B}\ }\textbf {\bibinfo {volume} {77}},\ \bibinfo
  {pages} {104445} (\bibinfo {year} {2008})}\BibitemShut {NoStop}%
\bibitem [{\citenamefont {Liu}\ \emph {et~al.}(2009)\citenamefont {Liu},
  \citenamefont {Obi}, \citenamefont {Lou}, \citenamefont {Chen}, \citenamefont
  {Cai}, \citenamefont {Stoute}, \citenamefont {Espanol}, \citenamefont {Lew},
  \citenamefont {Situ}, \citenamefont {Ziemer}, \citenamefont {Harris},\ and\
  \citenamefont {Sun}}]{Liu:19:2009}%
  \BibitemOpen
  \bibfield  {author} {\bibinfo {author} {\bibfnamefont {M.}~\bibnamefont
  {Liu}}, \bibinfo {author} {\bibfnamefont {O.}~\bibnamefont {Obi}}, \bibinfo
  {author} {\bibfnamefont {J.}~\bibnamefont {Lou}}, \bibinfo {author}
  {\bibfnamefont {Y.}~\bibnamefont {Chen}}, \bibinfo {author} {\bibfnamefont
  {Z.}~\bibnamefont {Cai}}, \bibinfo {author} {\bibfnamefont {S.}~\bibnamefont
  {Stoute}}, \bibinfo {author} {\bibfnamefont {M.}~\bibnamefont {Espanol}},
  \bibinfo {author} {\bibfnamefont {M.}~\bibnamefont {Lew}}, \bibinfo {author}
  {\bibfnamefont {X.}~\bibnamefont {Situ}}, \bibinfo {author} {\bibfnamefont
  {K.~S.}\ \bibnamefont {Ziemer}}, \bibinfo {author} {\bibfnamefont {V.~G.}\
  \bibnamefont {Harris}}, \ and\ \bibinfo {author} {\bibfnamefont {N.~X.}\
  \bibnamefont {Sun}},\ }\href {\doibase 10.1002/adfm.200801907} {\bibfield
  {journal} {\bibinfo  {journal} {Adv. Funct. Mater.}\ }\textbf {\bibinfo
  {volume} {19}},\ \bibinfo {pages} {1826} (\bibinfo {year}
  {2009})}\BibitemShut {NoStop}%
\bibitem [{\citenamefont {Liu}\ \emph {et~al.}(2010)\citenamefont {Liu},
  \citenamefont {Obi}, \citenamefont {Cai}, \citenamefont {Lou}, \citenamefont
  {Yang}, \citenamefont {Ziemer},\ and\ \citenamefont {Sun}}]{Liu:107:2010}%
  \BibitemOpen
  \bibfield  {author} {\bibinfo {author} {\bibfnamefont {M.}~\bibnamefont
  {Liu}}, \bibinfo {author} {\bibfnamefont {O.}~\bibnamefont {Obi}}, \bibinfo
  {author} {\bibfnamefont {Z.}~\bibnamefont {Cai}}, \bibinfo {author}
  {\bibfnamefont {J.}~\bibnamefont {Lou}}, \bibinfo {author} {\bibfnamefont
  {G.}~\bibnamefont {Yang}}, \bibinfo {author} {\bibfnamefont {K.~S.}\
  \bibnamefont {Ziemer}}, \ and\ \bibinfo {author} {\bibfnamefont {N.~X.}\
  \bibnamefont {Sun}},\ }\href {\doibase 10.1063/1.3354104} {\bibfield
  {journal} {\bibinfo  {journal} {J. Appl. Phys.}\ }\textbf {\bibinfo {volume}
  {107}},\ \bibinfo {pages} {073916} (\bibinfo {year} {2010})}\BibitemShut
  {NoStop}%
\bibitem [{\citenamefont {O'Handley}(2000)}]{OHandley:book}%
  \BibitemOpen
  \bibfield  {author} {\bibinfo {author} {\bibfnamefont {R.~C.}\ \bibnamefont
  {O'Handley}},\ }\href@noop {} {\emph {\bibinfo {title} {Modern Magnetic
  Materials: Principles and Applications}}},\ \bibinfo {edition} {1st}\ ed.\
  (\bibinfo  {publisher} {John Wiley \& Sons},\ \bibinfo {year}
  {2000})\BibitemShut {NoStop}%
\bibitem [{\citenamefont {Arai}\ \emph {et~al.}(1976)\citenamefont {Arai},
  \citenamefont {Ohmori}, \citenamefont {Tsuya},\ and\ \citenamefont
  {Iida}}]{Arai:34:1976}%
  \BibitemOpen
  \bibfield  {author} {\bibinfo {author} {\bibfnamefont {K.~I.}\ \bibnamefont
  {Arai}}, \bibinfo {author} {\bibfnamefont {K.}~\bibnamefont {Ohmori}},
  \bibinfo {author} {\bibfnamefont {N.}~\bibnamefont {Tsuya}}, \ and\ \bibinfo
  {author} {\bibfnamefont {S.}~\bibnamefont {Iida}},\ }\href {\doibase
  10.1002/pssa.2210340129} {\bibfield  {journal} {\bibinfo  {journal} {Phys.
  Status Solidi A}\ }\textbf {\bibinfo {volume} {34}},\ \bibinfo {pages} {325}
  (\bibinfo {year} {1976})}\BibitemShut {NoStop}%
\bibitem [{\citenamefont {Zhang}\ and\ \citenamefont
  {Satpathy}(1991)}]{Zhang:44:1991}%
  \BibitemOpen
  \bibfield  {author} {\bibinfo {author} {\bibfnamefont {Z.}~\bibnamefont
  {Zhang}}\ and\ \bibinfo {author} {\bibfnamefont {S.}~\bibnamefont
  {Satpathy}},\ }\href {\doibase 10.1103/PhysRevB.44.13319} {\bibfield
  {journal} {\bibinfo  {journal} {Phys. Rev. B}\ }\textbf {\bibinfo {volume}
  {44}},\ \bibinfo {pages} {13319} (\bibinfo {year} {1991})}\BibitemShut
  {NoStop}%
\bibitem [{\citenamefont {Alexe}\ \emph {et~al.}(2009)\citenamefont {Alexe},
  \citenamefont {Ziese}, \citenamefont {Hesse}, \citenamefont {Esquinazi},
  \citenamefont {Yamauchi}, \citenamefont {Fukushima}, \citenamefont
  {Picozzi},\ and\ \citenamefont {G\"osele}}]{Alexe:21:2009}%
  \BibitemOpen
  \bibfield  {author} {\bibinfo {author} {\bibfnamefont {M.}~\bibnamefont
  {Alexe}}, \bibinfo {author} {\bibfnamefont {M.}~\bibnamefont {Ziese}},
  \bibinfo {author} {\bibfnamefont {D.}~\bibnamefont {Hesse}}, \bibinfo
  {author} {\bibfnamefont {P.}~\bibnamefont {Esquinazi}}, \bibinfo {author}
  {\bibfnamefont {K.}~\bibnamefont {Yamauchi}}, \bibinfo {author}
  {\bibfnamefont {T.}~\bibnamefont {Fukushima}}, \bibinfo {author}
  {\bibfnamefont {S.}~\bibnamefont {Picozzi}}, \ and\ \bibinfo {author}
  {\bibfnamefont {U.}~\bibnamefont {G\"osele}},\ }\href {\doibase
  10.1002/adma.200901381} {\bibfield  {journal} {\bibinfo  {journal} {Adv.
  Mater.}\ }\textbf {\bibinfo {volume} {21}},\ \bibinfo {pages} {4452}
  (\bibinfo {year} {2009})}\BibitemShut {NoStop}%
\bibitem [{\citenamefont {Takahashi}\ \emph {et~al.}(2012)\citenamefont
  {Takahashi}, \citenamefont {Misumi},\ and\ \citenamefont
  {Lippmaa}}]{Takahashi:86:2012}%
  \BibitemOpen
  \bibfield  {author} {\bibinfo {author} {\bibfnamefont {R.}~\bibnamefont
  {Takahashi}}, \bibinfo {author} {\bibfnamefont {H.}~\bibnamefont {Misumi}}, \
  and\ \bibinfo {author} {\bibfnamefont {M.}~\bibnamefont {Lippmaa}},\ }\href
  {\doibase 10.1103/PhysRevB.86.144105} {\bibfield  {journal} {\bibinfo
  {journal} {Phys. Rev. B}\ }\textbf {\bibinfo {volume} {86}},\ \bibinfo
  {pages} {144105} (\bibinfo {year} {2012})}\BibitemShut {NoStop}%
\bibitem [{\citenamefont {Nazarenko}\ \emph {et~al.}(2006)\citenamefont
  {Nazarenko}, \citenamefont {Lorenzo}, \citenamefont {Joly}, \citenamefont
  {Hodeau}, \citenamefont {Mannix},\ and\ \citenamefont
  {Marin}}]{Nazarenko:97:2006}%
  \BibitemOpen
  \bibfield  {author} {\bibinfo {author} {\bibfnamefont {E.}~\bibnamefont
  {Nazarenko}}, \bibinfo {author} {\bibfnamefont {J.~E.}\ \bibnamefont
  {Lorenzo}}, \bibinfo {author} {\bibfnamefont {Y.}~\bibnamefont {Joly}},
  \bibinfo {author} {\bibfnamefont {J.~L.}\ \bibnamefont {Hodeau}}, \bibinfo
  {author} {\bibfnamefont {D.}~\bibnamefont {Mannix}}, \ and\ \bibinfo {author}
  {\bibfnamefont {C.}~\bibnamefont {Marin}},\ }\href {\doibase
  10.1103/PhysRevLett.97.056403} {\bibfield  {journal} {\bibinfo  {journal}
  {Phys. Rev. Lett.}\ }\textbf {\bibinfo {volume} {97}},\ \bibinfo {pages}
  {056403} (\bibinfo {year} {2006})}\BibitemShut {NoStop}%
\bibitem [{\citenamefont {Yamauchi}\ \emph {et~al.}(2009)\citenamefont
  {Yamauchi}, \citenamefont {Fukushima},\ and\ \citenamefont
  {Picozzi}}]{Yamauchi:79:2009}%
  \BibitemOpen
  \bibfield  {author} {\bibinfo {author} {\bibfnamefont {K.}~\bibnamefont
  {Yamauchi}}, \bibinfo {author} {\bibfnamefont {T.}~\bibnamefont {Fukushima}},
  \ and\ \bibinfo {author} {\bibfnamefont {S.}~\bibnamefont {Picozzi}},\ }\href
  {\doibase 10.1103/PhysRevB.79.212404} {\bibfield  {journal} {\bibinfo
  {journal} {Phys. Rev. B}\ }\textbf {\bibinfo {volume} {79}},\ \bibinfo
  {pages} {212404} (\bibinfo {year} {2009})}\BibitemShut {NoStop}%
\bibitem [{\citenamefont {Rado}\ and\ \citenamefont
  {Ferrari}(1975)}]{Rado:12:1975}%
  \BibitemOpen
  \bibfield  {author} {\bibinfo {author} {\bibfnamefont {G.~T.}\ \bibnamefont
  {Rado}}\ and\ \bibinfo {author} {\bibfnamefont {J.~M.}\ \bibnamefont
  {Ferrari}},\ }\href {\doibase 10.1103/PhysRevB.12.5166} {\bibfield  {journal}
  {\bibinfo  {journal} {Phys. Rev. B}\ }\textbf {\bibinfo {volume} {12}},\
  \bibinfo {pages} {5166} (\bibinfo {year} {1975})}\BibitemShut {NoStop}%
\bibitem [{\citenamefont {Gepr\"ags}\ \emph {et~al.}(2012)\citenamefont
  {Gepr\"ags}, \citenamefont {Opel}, \citenamefont {Goennenwein},\ and\
  \citenamefont {Gross}}]{Gepraegs:86:2012}%
  \BibitemOpen
  \bibfield  {author} {\bibinfo {author} {\bibfnamefont {S.}~\bibnamefont
  {Gepr\"ags}}, \bibinfo {author} {\bibfnamefont {M.}~\bibnamefont {Opel}},
  \bibinfo {author} {\bibfnamefont {S.~T.~B.}\ \bibnamefont {Goennenwein}}, \
  and\ \bibinfo {author} {\bibfnamefont {R.}~\bibnamefont {Gross}},\ }\href
  {\doibase 10.1103/PhysRevB.86.134432} {\bibfield  {journal} {\bibinfo
  {journal} {Phys. Rev. B}\ }\textbf {\bibinfo {volume} {86}},\ \bibinfo
  {pages} {134432} (\bibinfo {year} {2012})}\BibitemShut {NoStop}%
\bibitem [{\citenamefont {Gross}\ \emph {et~al.}(2000)\citenamefont {Gross},
  \citenamefont {Klein}, \citenamefont {Wiedenhorst}, \citenamefont
  {H\"ofener}, \citenamefont {Schoop}, \citenamefont {Philipp}, \citenamefont
  {Schonecke}, \citenamefont {Herbstritt}, \citenamefont {Alff}, \citenamefont
  {Lu}, \citenamefont {Marx}, \citenamefont {Schymon}, \citenamefont
  {Thienhaus},\ and\ \citenamefont {Mader}}]{Gross:4058:2000}%
  \BibitemOpen
  \bibfield  {author} {\bibinfo {author} {\bibfnamefont {R.}~\bibnamefont
  {Gross}}, \bibinfo {author} {\bibfnamefont {J.}~\bibnamefont {Klein}},
  \bibinfo {author} {\bibfnamefont {B.}~\bibnamefont {Wiedenhorst}}, \bibinfo
  {author} {\bibfnamefont {C.}~\bibnamefont {H\"ofener}}, \bibinfo {author}
  {\bibfnamefont {U.}~\bibnamefont {Schoop}}, \bibinfo {author} {\bibfnamefont
  {J.~B.}\ \bibnamefont {Philipp}}, \bibinfo {author} {\bibfnamefont
  {M.}~\bibnamefont {Schonecke}}, \bibinfo {author} {\bibfnamefont
  {F.}~\bibnamefont {Herbstritt}}, \bibinfo {author} {\bibfnamefont
  {L.}~\bibnamefont {Alff}}, \bibinfo {author} {\bibfnamefont {Y.}~\bibnamefont
  {Lu}}, \bibinfo {author} {\bibfnamefont {A.}~\bibnamefont {Marx}}, \bibinfo
  {author} {\bibfnamefont {S.}~\bibnamefont {Schymon}}, \bibinfo {author}
  {\bibfnamefont {S.}~\bibnamefont {Thienhaus}}, \ and\ \bibinfo {author}
  {\bibfnamefont {W.}~\bibnamefont {Mader}},\ }\href {\doibase
  10.1117/12.397845} {\bibfield  {journal} {\bibinfo  {journal} {Proc. SPIE}\
  }\textbf {\bibinfo {volume} {4058}},\ \bibinfo {pages} {278} (\bibinfo {year}
  {2000})}\BibitemShut {NoStop}%
\bibitem [{\citenamefont {Klein}\ \emph {et~al.}(1999)\citenamefont {Klein},
  \citenamefont {H{\"o}fener}, \citenamefont {Alff},\ and\ \citenamefont
  {Gross}}]{Klein:12:1999}%
  \BibitemOpen
  \bibfield  {author} {\bibinfo {author} {\bibfnamefont {J.}~\bibnamefont
  {Klein}}, \bibinfo {author} {\bibfnamefont {C.}~\bibnamefont {H{\"o}fener}},
  \bibinfo {author} {\bibfnamefont {L.}~\bibnamefont {Alff}}, \ and\ \bibinfo
  {author} {\bibfnamefont {R.}~\bibnamefont {Gross}},\ }\href {\doibase
  10.1088/0953-2048/12/11/398} {\bibfield  {journal} {\bibinfo  {journal}
  {Supercond. Sci. Technol.}\ }\textbf {\bibinfo {volume} {12}},\ \bibinfo
  {pages} {1023} (\bibinfo {year} {1999})}\BibitemShut {NoStop}%
\bibitem [{\citenamefont {Klein}\ \emph {et~al.}(2000)\citenamefont {Klein},
  \citenamefont {H\"ofener}, \citenamefont {Alff},\ and\ \citenamefont
  {Gross}}]{Klein:211:2000}%
  \BibitemOpen
  \bibfield  {author} {\bibinfo {author} {\bibfnamefont {J.}~\bibnamefont
  {Klein}}, \bibinfo {author} {\bibfnamefont {C.}~\bibnamefont {H\"ofener}},
  \bibinfo {author} {\bibfnamefont {L.}~\bibnamefont {Alff}}, \ and\ \bibinfo
  {author} {\bibfnamefont {R.}~\bibnamefont {Gross}},\ }\href {\doibase
  10.1016/S0304-8853(99)00706-4} {\bibfield  {journal} {\bibinfo  {journal} {J.
  Magn. Magn. Mater.}\ }\textbf {\bibinfo {volume} {211}},\ \bibinfo {pages}
  {9} (\bibinfo {year} {2000})}\BibitemShut {NoStop}%
\bibitem [{\citenamefont {Reisinger}\ \emph {et~al.}(2003)\citenamefont
  {Reisinger}, \citenamefont {Blass}, \citenamefont {Klein}, \citenamefont
  {Philipp}, \citenamefont {Schonecke}, \citenamefont {Erb}, \citenamefont
  {Alff},\ and\ \citenamefont {Gross}}]{Reisinger:77:2003}%
  \BibitemOpen
  \bibfield  {author} {\bibinfo {author} {\bibfnamefont {D.}~\bibnamefont
  {Reisinger}}, \bibinfo {author} {\bibfnamefont {B.}~\bibnamefont {Blass}},
  \bibinfo {author} {\bibfnamefont {J.}~\bibnamefont {Klein}}, \bibinfo
  {author} {\bibfnamefont {J.}~\bibnamefont {Philipp}}, \bibinfo {author}
  {\bibfnamefont {M.}~\bibnamefont {Schonecke}}, \bibinfo {author}
  {\bibfnamefont {A.}~\bibnamefont {Erb}}, \bibinfo {author} {\bibfnamefont
  {L.}~\bibnamefont {Alff}}, \ and\ \bibinfo {author} {\bibfnamefont
  {R.}~\bibnamefont {Gross}},\ }\href {\doibase 10.1007/s00339-003-2105-9}
  {\bibfield  {journal} {\bibinfo  {journal} {Appl. Phys. A}\ }\textbf
  {\bibinfo {volume} {77}},\ \bibinfo {pages} {619} (\bibinfo {year}
  {2003})}\BibitemShut {NoStop}%
\bibitem [{\citenamefont {Venkateshvaran}\ \emph {et~al.}(2009)\citenamefont
  {Venkateshvaran}, \citenamefont {Althammer}, \citenamefont {Nielsen},
  \citenamefont {Gepr\"ags}, \citenamefont {Ramachandra~Rao}, \citenamefont
  {Goennenwein}, \citenamefont {Opel},\ and\ \citenamefont
  {Gross}}]{Venkateshvaran:79:2009}%
  \BibitemOpen
  \bibfield  {author} {\bibinfo {author} {\bibfnamefont {D.}~\bibnamefont
  {Venkateshvaran}}, \bibinfo {author} {\bibfnamefont {M.}~\bibnamefont
  {Althammer}}, \bibinfo {author} {\bibfnamefont {A.}~\bibnamefont {Nielsen}},
  \bibinfo {author} {\bibfnamefont {S.}~\bibnamefont {Gepr\"ags}}, \bibinfo
  {author} {\bibfnamefont {M.~S.}\ \bibnamefont {Ramachandra~Rao}}, \bibinfo
  {author} {\bibfnamefont {S.~T.~B.}\ \bibnamefont {Goennenwein}}, \bibinfo
  {author} {\bibfnamefont {M.}~\bibnamefont {Opel}}, \ and\ \bibinfo {author}
  {\bibfnamefont {R.}~\bibnamefont {Gross}},\ }\href {\doibase
  10.1103/PhysRevB.79.134405} {\bibfield  {journal} {\bibinfo  {journal} {Phys.
  Rev. B}\ }\textbf {\bibinfo {volume} {79}},\ \bibinfo {pages} {134405}
  (\bibinfo {year} {2009})}\BibitemShut {NoStop}%
\bibitem [{\citenamefont {Reisinger}\ \emph {et~al.}(2004)\citenamefont
  {Reisinger}, \citenamefont {Majewski}, \citenamefont {Opel}, \citenamefont
  {Alff},\ and\ \citenamefont {Gross}}]{Reisinger:85:2004}%
  \BibitemOpen
  \bibfield  {author} {\bibinfo {author} {\bibfnamefont {D.}~\bibnamefont
  {Reisinger}}, \bibinfo {author} {\bibfnamefont {P.}~\bibnamefont {Majewski}},
  \bibinfo {author} {\bibfnamefont {M.}~\bibnamefont {Opel}}, \bibinfo {author}
  {\bibfnamefont {L.}~\bibnamefont {Alff}}, \ and\ \bibinfo {author}
  {\bibfnamefont {R.}~\bibnamefont {Gross}},\ }\href {\doibase
  10.1063/1.1808497} {\bibfield  {journal} {\bibinfo  {journal} {Appl. Phys.
  Lett.}\ }\textbf {\bibinfo {volume} {85}},\ \bibinfo {pages} {4980} (\bibinfo
  {year} {2004})}\BibitemShut {NoStop}%
\bibitem [{\citenamefont {Bataille}\ \emph {et~al.}(2006)\citenamefont
  {Bataille}, \citenamefont {Ponson}, \citenamefont {Gota}, \citenamefont
  {Barbier}, \citenamefont {Bonamy}, \citenamefont {Gautier-Soyer},
  \citenamefont {Gatel},\ and\ \citenamefont {Snoeck}}]{Bataille:74:2006}%
  \BibitemOpen
  \bibfield  {author} {\bibinfo {author} {\bibfnamefont {A.~M.}\ \bibnamefont
  {Bataille}}, \bibinfo {author} {\bibfnamefont {L.}~\bibnamefont {Ponson}},
  \bibinfo {author} {\bibfnamefont {S.}~\bibnamefont {Gota}}, \bibinfo {author}
  {\bibfnamefont {L.}~\bibnamefont {Barbier}}, \bibinfo {author} {\bibfnamefont
  {D.}~\bibnamefont {Bonamy}}, \bibinfo {author} {\bibfnamefont
  {M.}~\bibnamefont {Gautier-Soyer}}, \bibinfo {author} {\bibfnamefont
  {C.}~\bibnamefont {Gatel}}, \ and\ \bibinfo {author} {\bibfnamefont
  {E.}~\bibnamefont {Snoeck}},\ }\href {\doibase 10.1103/PhysRevB.74.155438}
  {\bibfield  {journal} {\bibinfo  {journal} {Phys. Rev. B}\ }\textbf {\bibinfo
  {volume} {74}},\ \bibinfo {pages} {155438} (\bibinfo {year}
  {2006})}\BibitemShut {NoStop}%
\bibitem [{\citenamefont {Sofin}\ \emph {et~al.}(2011)\citenamefont {Sofin},
  \citenamefont {Arora},\ and\ \citenamefont {Shvets}}]{Sofin:83:2011}%
  \BibitemOpen
  \bibfield  {author} {\bibinfo {author} {\bibfnamefont {R.~G.~S.}\
  \bibnamefont {Sofin}}, \bibinfo {author} {\bibfnamefont {S.~K.}\ \bibnamefont
  {Arora}}, \ and\ \bibinfo {author} {\bibfnamefont {I.~V.}\ \bibnamefont
  {Shvets}},\ }\href {\doibase 10.1103/PhysRevB.83.134436} {\bibfield
  {journal} {\bibinfo  {journal} {Phys. Rev. B}\ }\textbf {\bibinfo {volume}
  {83}},\ \bibinfo {pages} {134436} (\bibinfo {year} {2011})}\BibitemShut
  {NoStop}%
\bibitem [{\citenamefont {Kakol}\ and\ \citenamefont
  {Honig}(1989)}]{Kakol:40:1989}%
  \BibitemOpen
  \bibfield  {author} {\bibinfo {author} {\bibfnamefont {Z.}~\bibnamefont
  {Kakol}}\ and\ \bibinfo {author} {\bibfnamefont {J.~M.}\ \bibnamefont
  {Honig}},\ }\href {\doibase 10.1103/PhysRevB.40.9090} {\bibfield  {journal}
  {\bibinfo  {journal} {Phys. Rev. B}\ }\textbf {\bibinfo {volume} {40}},\
  \bibinfo {pages} {9090} (\bibinfo {year} {1989})}\BibitemShut {NoStop}%
\bibitem [{\citenamefont {Shepherd}\ \emph {et~al.}(1991)\citenamefont
  {Shepherd}, \citenamefont {Koenitzer}, \citenamefont {Arag\'on},
  \citenamefont {Spalek},\ and\ \citenamefont {Honig}}]{Shepherd:43:1991}%
  \BibitemOpen
  \bibfield  {author} {\bibinfo {author} {\bibfnamefont {J.~P.}\ \bibnamefont
  {Shepherd}}, \bibinfo {author} {\bibfnamefont {J.~W.}\ \bibnamefont
  {Koenitzer}}, \bibinfo {author} {\bibfnamefont {R.}~\bibnamefont {Arag\'on}},
  \bibinfo {author} {\bibfnamefont {J.}~\bibnamefont {Spalek}}, \ and\ \bibinfo
  {author} {\bibfnamefont {J.~M.}\ \bibnamefont {Honig}},\ }\href {\doibase
  10.1103/PhysRevB.43.8461} {\bibfield  {journal} {\bibinfo  {journal} {Phys.
  Rev. B}\ }\textbf {\bibinfo {volume} {43}},\ \bibinfo {pages} {8461}
  (\bibinfo {year} {1991})}\BibitemShut {NoStop}%
\bibitem [{\citenamefont {Astrov}(1960)}]{Astrov:11:1960}%
  \BibitemOpen
  \bibfield  {author} {\bibinfo {author} {\bibfnamefont {D.~N.}\ \bibnamefont
  {Astrov}},\ }\href@noop {} {\bibfield  {journal} {\bibinfo  {journal} {JETP}\
  }\textbf {\bibinfo {volume} {11}},\ \bibinfo {pages} {708} (\bibinfo {year}
  {1960})}\BibitemShut {NoStop}%
\bibitem [{\citenamefont {Kitagawa}\ \emph {et~al.}(2010)\citenamefont
  {Kitagawa}, \citenamefont {Hiraoka}, \citenamefont {Honda}, \citenamefont
  {Ishikura}, \citenamefont {Nakamura},\ and\ \citenamefont
  {Kimura}}]{Kitagawa:9:2010}%
  \BibitemOpen
  \bibfield  {author} {\bibinfo {author} {\bibfnamefont {Y.}~\bibnamefont
  {Kitagawa}}, \bibinfo {author} {\bibfnamefont {Y.}~\bibnamefont {Hiraoka}},
  \bibinfo {author} {\bibfnamefont {T.}~\bibnamefont {Honda}}, \bibinfo
  {author} {\bibfnamefont {T.}~\bibnamefont {Ishikura}}, \bibinfo {author}
  {\bibfnamefont {H.}~\bibnamefont {Nakamura}}, \ and\ \bibinfo {author}
  {\bibfnamefont {T.}~\bibnamefont {Kimura}},\ }\href {\doibase
  10.1038/nmat2826} {\bibfield  {journal} {\bibinfo  {journal} {Nat. Mater.}\
  }\textbf {\bibinfo {volume} {9}},\ \bibinfo {pages} {797} (\bibinfo {year}
  {2010})}\BibitemShut {NoStop}%
\bibitem [{\citenamefont {Chun}\ \emph {et~al.}(2012)\citenamefont {Chun},
  \citenamefont {Chai}, \citenamefont {Jeon}, \citenamefont {Kim},
  \citenamefont {Oh}, \citenamefont {Kim}, \citenamefont {Kim}, \citenamefont
  {Jeon}, \citenamefont {Haam}, \citenamefont {Park}, \citenamefont {Lee},
  \citenamefont {Chung}, \citenamefont {Park},\ and\ \citenamefont
  {Kim}}]{Chun:108:2012}%
  \BibitemOpen
  \bibfield  {author} {\bibinfo {author} {\bibfnamefont {S.~H.}\ \bibnamefont
  {Chun}}, \bibinfo {author} {\bibfnamefont {Y.~S.}\ \bibnamefont {Chai}},
  \bibinfo {author} {\bibfnamefont {B.-G.}\ \bibnamefont {Jeon}}, \bibinfo
  {author} {\bibfnamefont {H.~J.}\ \bibnamefont {Kim}}, \bibinfo {author}
  {\bibfnamefont {Y.~S.}\ \bibnamefont {Oh}}, \bibinfo {author} {\bibfnamefont
  {I.}~\bibnamefont {Kim}}, \bibinfo {author} {\bibfnamefont {H.}~\bibnamefont
  {Kim}}, \bibinfo {author} {\bibfnamefont {B.~J.}\ \bibnamefont {Jeon}},
  \bibinfo {author} {\bibfnamefont {S.~Y.}\ \bibnamefont {Haam}}, \bibinfo
  {author} {\bibfnamefont {J.-Y.}\ \bibnamefont {Park}}, \bibinfo {author}
  {\bibfnamefont {S.~H.}\ \bibnamefont {Lee}}, \bibinfo {author} {\bibfnamefont
  {J.-H.}\ \bibnamefont {Chung}}, \bibinfo {author} {\bibfnamefont {J.-H.}\
  \bibnamefont {Park}}, \ and\ \bibinfo {author} {\bibfnamefont {K.~H.}\
  \bibnamefont {Kim}},\ }\href {\doibase 10.1103/PhysRevLett.108.177201}
  {\bibfield  {journal} {\bibinfo  {journal} {Phys. Rev. Lett.}\ }\textbf
  {\bibinfo {volume} {108}},\ \bibinfo {pages} {177201} (\bibinfo {year}
  {2012})}\BibitemShut {NoStop}%
\bibitem [{\citenamefont {Hall}\ \emph {et~al.}(1977)\citenamefont {Hall},
  \citenamefont {Veeraraghavan}, \citenamefont {Rubin},\ and\ \citenamefont
  {Winchell}}]{Hall:10:1977}%
  \BibitemOpen
  \bibfield  {author} {\bibinfo {author} {\bibfnamefont {M.~M.}\ \bibnamefont
  {Hall}}, \bibinfo {author} {\bibfnamefont {V.~G.}\ \bibnamefont
  {Veeraraghavan}}, \bibinfo {author} {\bibfnamefont {H.}~\bibnamefont
  {Rubin}}, \ and\ \bibinfo {author} {\bibfnamefont {P.~G.}\ \bibnamefont
  {Winchell}},\ }\href {\doibase 10.1107/S0021889877012849} {\bibfield
  {journal} {\bibinfo  {journal} {J. Appl. Crystallogr.}\ }\textbf {\bibinfo
  {volume} {10}},\ \bibinfo {pages} {66} (\bibinfo {year} {1977})}\BibitemShut
  {NoStop}%
\bibitem [{\citenamefont {Schwenk}\ \emph {et~al.}(2000)\citenamefont
  {Schwenk}, \citenamefont {Bareiter}, \citenamefont {Hinkel}, \citenamefont
  {L\"{u}thi}, \citenamefont {Kakol}, \citenamefont {Koslowski},\ and\
  \citenamefont {Honig}}]{Schwenk:13:2000}%
  \BibitemOpen
  \bibfield  {author} {\bibinfo {author} {\bibfnamefont {H.}~\bibnamefont
  {Schwenk}}, \bibinfo {author} {\bibfnamefont {S.}~\bibnamefont {Bareiter}},
  \bibinfo {author} {\bibfnamefont {C.}~\bibnamefont {Hinkel}}, \bibinfo
  {author} {\bibfnamefont {B.}~\bibnamefont {L\"{u}thi}}, \bibinfo {author}
  {\bibfnamefont {Z.}~\bibnamefont {Kakol}}, \bibinfo {author} {\bibfnamefont
  {A.}~\bibnamefont {Koslowski}}, \ and\ \bibinfo {author} {\bibfnamefont
  {J.~M.}\ \bibnamefont {Honig}},\ }\href {\doibase 10.1007/s100510050060}
  {\bibfield  {journal} {\bibinfo  {journal} {Eur. Phys. J. B}\ }\textbf
  {\bibinfo {volume} {13}},\ \bibinfo {pages} {491} (\bibinfo {year}
  {2000})}\BibitemShut {NoStop}%
\bibitem [{\citenamefont {Fu}\ and\ \citenamefont {Cohen}(2000)}]{Fu:403:2000}%
  \BibitemOpen
  \bibfield  {author} {\bibinfo {author} {\bibfnamefont {H.}~\bibnamefont
  {Fu}}\ and\ \bibinfo {author} {\bibfnamefont {R.~E.}\ \bibnamefont {Cohen}},\
  }\href {\doibase 10.1038/35002022} {\bibfield  {journal} {\bibinfo  {journal}
  {Nature}\ }\textbf {\bibinfo {volume} {403}},\ \bibinfo {pages} {281}
  (\bibinfo {year} {2000})}\BibitemShut {NoStop}%
\bibitem [{\citenamefont {Fong}\ and\ \citenamefont
  {Thompson}(2006)}]{Fong:36:2006}%
  \BibitemOpen
  \bibfield  {author} {\bibinfo {author} {\bibfnamefont {D.~D.}\ \bibnamefont
  {Fong}}\ and\ \bibinfo {author} {\bibfnamefont {C.}~\bibnamefont
  {Thompson}},\ }\href {\doibase 10.1146/annurev.matsci.36.090804.100242}
  {\bibfield  {journal} {\bibinfo  {journal} {Ann. Rev. Mater. Res.}\ }\textbf
  {\bibinfo {volume} {36}},\ \bibinfo {pages} {431} (\bibinfo {year}
  {2006})}\BibitemShut {NoStop}%
\bibitem [{\citenamefont {van Reeuwijk}\ \emph {et~al.}(2004)\citenamefont {van
  Reeuwijk}, \citenamefont {Karakaya}, \citenamefont {Graafsma},\ and\
  \citenamefont {Harkema}}]{Reeuwijk:37:2004}%
  \BibitemOpen
  \bibfield  {author} {\bibinfo {author} {\bibfnamefont {S.~J.}\ \bibnamefont
  {van Reeuwijk}}, \bibinfo {author} {\bibfnamefont {K.}~\bibnamefont
  {Karakaya}}, \bibinfo {author} {\bibfnamefont {H.}~\bibnamefont {Graafsma}},
  \ and\ \bibinfo {author} {\bibfnamefont {S.}~\bibnamefont {Harkema}},\ }\href
  {\doibase 10.1107/S0021889803028395} {\bibfield  {journal} {\bibinfo
  {journal} {J. Appl. Crystallogr.}\ }\textbf {\bibinfo {volume} {37}},\
  \bibinfo {pages} {193} (\bibinfo {year} {2004})}\BibitemShut {NoStop}%
\bibitem [{\citenamefont {Subbarao}\ \emph {et~al.}(1957)\citenamefont
  {Subbarao}, \citenamefont {McQuarrie},\ and\ \citenamefont
  {Buessem}}]{Subbarao:28:1957}%
  \BibitemOpen
  \bibfield  {author} {\bibinfo {author} {\bibfnamefont {E.~C.}\ \bibnamefont
  {Subbarao}}, \bibinfo {author} {\bibfnamefont {M.~C.}\ \bibnamefont
  {McQuarrie}}, \ and\ \bibinfo {author} {\bibfnamefont {W.~R.}\ \bibnamefont
  {Buessem}},\ }\href {\doibase 10.1063/1.1722606} {\bibfield  {journal}
  {\bibinfo  {journal} {J. Appl. Phys.}\ }\textbf {\bibinfo {volume} {28}},\
  \bibinfo {pages} {1194} (\bibinfo {year} {1957})}\BibitemShut {NoStop}%
\bibitem [{\citenamefont {Lengsdorf}\ \emph {et~al.}(2004)\citenamefont
  {Lengsdorf}, \citenamefont {Ait-Tahar}, \citenamefont {Saxena}, \citenamefont
  {Ellerby}, \citenamefont {Khomskii}, \citenamefont {Micklitz}, \citenamefont
  {Lorenz},\ and\ \citenamefont {Abd-Elmeguid}}]{Lengsdorf:69:2004}%
  \BibitemOpen
  \bibfield  {author} {\bibinfo {author} {\bibfnamefont {R.}~\bibnamefont
  {Lengsdorf}}, \bibinfo {author} {\bibfnamefont {M.}~\bibnamefont
  {Ait-Tahar}}, \bibinfo {author} {\bibfnamefont {S.~S.}\ \bibnamefont
  {Saxena}}, \bibinfo {author} {\bibfnamefont {M.}~\bibnamefont {Ellerby}},
  \bibinfo {author} {\bibfnamefont {D.~I.}\ \bibnamefont {Khomskii}}, \bibinfo
  {author} {\bibfnamefont {H.}~\bibnamefont {Micklitz}}, \bibinfo {author}
  {\bibfnamefont {T.}~\bibnamefont {Lorenz}}, \ and\ \bibinfo {author}
  {\bibfnamefont {M.~M.}\ \bibnamefont {Abd-Elmeguid}},\ }\href {\doibase
  10.1103/PhysRevB.69.140403} {\bibfield  {journal} {\bibinfo  {journal} {Phys.
  Rev. B}\ }\textbf {\bibinfo {volume} {69}},\ \bibinfo {pages} {140403}
  (\bibinfo {year} {2004})}\BibitemShut {NoStop}%
\bibitem [{\citenamefont {Nye}(1985)}]{Nye:book}%
  \BibitemOpen
  \bibfield  {author} {\bibinfo {author} {\bibfnamefont {J.}~\bibnamefont
  {Nye}},\ }\href@noop {} {\emph {\bibinfo {title} {Physical Properties of
  Crystals}}}\ (\bibinfo  {publisher} {Oxford University Press},\ \bibinfo
  {year} {1985})\BibitemShut {NoStop}%
\bibitem [{\citenamefont {Tian}\ \emph {et~al.}(2009)\citenamefont {Tian},
  \citenamefont {Sander},\ and\ \citenamefont {Kirschner}}]{Tian:79:2009}%
  \BibitemOpen
  \bibfield  {author} {\bibinfo {author} {\bibfnamefont {Z.}~\bibnamefont
  {Tian}}, \bibinfo {author} {\bibfnamefont {D.}~\bibnamefont {Sander}}, \ and\
  \bibinfo {author} {\bibfnamefont {J.}~\bibnamefont {Kirschner}},\ }\href
  {\doibase 10.1103/PhysRevB.79.024432} {\bibfield  {journal} {\bibinfo
  {journal} {Phys. Rev. B}\ }\textbf {\bibinfo {volume} {79}},\ \bibinfo
  {pages} {024432} (\bibinfo {year} {2009})}\BibitemShut {NoStop}%
\bibitem [{\citenamefont {Sun}\ and\ \citenamefont
  {O'Handley}(1991)}]{Sun:66:1991}%
  \BibitemOpen
  \bibfield  {author} {\bibinfo {author} {\bibfnamefont {S.~W.}\ \bibnamefont
  {Sun}}\ and\ \bibinfo {author} {\bibfnamefont {R.~C.}\ \bibnamefont
  {O'Handley}},\ }\href {\doibase 10.1103/PhysRevLett.66.2798} {\bibfield
  {journal} {\bibinfo  {journal} {Phys. Rev. Lett.}\ }\textbf {\bibinfo
  {volume} {66}},\ \bibinfo {pages} {2798} (\bibinfo {year}
  {1991})}\BibitemShut {NoStop}%
\bibitem [{\citenamefont {Bickford}\ \emph {et~al.}(1955)\citenamefont
  {Bickford}, \citenamefont {Pappis},\ and\ \citenamefont
  {Stull}}]{Bickford:99:1210}%
  \BibitemOpen
  \bibfield  {author} {\bibinfo {author} {\bibfnamefont {L.~R.}\ \bibnamefont
  {Bickford}}, \bibinfo {author} {\bibfnamefont {J.}~\bibnamefont {Pappis}}, \
  and\ \bibinfo {author} {\bibfnamefont {J.~L.}\ \bibnamefont {Stull}},\ }\href
  {\doibase 10.1103/PhysRev.99.1210} {\bibfield  {journal} {\bibinfo  {journal}
  {Phys. Rev.}\ }\textbf {\bibinfo {volume} {99}},\ \bibinfo {pages} {1210}
  (\bibinfo {year} {1955})}\BibitemShut {NoStop}%
\end{thebibliography}
\end{document}